\documentclass[aps,prd,preprint,floatfix,superscriptaddress,nofootinbib,longbibliography,a4paper]{revtex4-1}
\pdfoutput=1
\usepackage{color}
\usepackage{graphicx}
\usepackage{textcomp}
\usepackage{subfigure}
\usepackage{feynmp}
\usepackage{amsmath,amssymb,array}
\allowdisplaybreaks
\usepackage{enumerate}
\usepackage{hhline}
\usepackage{hyperref}
\usepackage{orcidlink}

\hypersetup{
    citecolor= red,
    colorlinks=true,
    linkcolor=blue,
    filecolor=magenta,      
    urlcolor=blue,}

\usepackage{epsfig}
\usepackage{xfrac}
\usepackage{slashed}
\usepackage{pifont}

\newcommand{\xmark}{\ding{55}}

\newcommand{\beqa}{\begin{eqnarray}}
\newcommand{\eeqa}{\end{eqnarray}}
\newcommand{\be}{\begin{equation}}
\newcommand{\ee}{\end{equation}}
\newcommand{\ba}{\begin{array}} 
\newcommand{\ea}{\end{array}}

\newcommand{\nl}{\nonumber \\}
\newcommand{\cc}{\, {\mathcal{C}}^{-1}\,}
\newcommand{\hc}{\;+\; \rm{H.c.}}
\newcommand{\so}{$SO(10)$}
\newcommand{\lag}{\mathcal{L}}
\newcommand{\fs}{\mathbf{16}}
\newcommand{\ft}{\mathbf{10}}
\newcommand{\ff}{\mathbf{\overline{5}}}
\newcommand{\fn}{\mathbf{1}}

\newcommand{\mh}{M^2_{H}}

\newcommand{\md}{M^2_{\Delta}}

\begin{document} 
\vspace*{0.5cm}
\title{Constraining scalars of $16_H$ through proton decays in non-renormalisable $SO(10)$ models}
\bigskip
%\author{Ketan M. Patel}
%\email{ketan.hep@gmail.com}
%\affiliation{Theoretical Physics Division, Physical Research Laboratory, Navarangpura, Ahmedabad-380009, India}
\author{Saurabh K. Shukla \orcidlink{0000-0001-5344-9889}}
\email{saurabhks@prl.res.in}
\affiliation{Theoretical Physics Division, Physical Research Laboratory, Navarangpura, Ahmedabad-380009, India}
\affiliation{Indian Institute of Technology Gandhinagar, Palaj-382055, India\vspace*{1cm}}

%----------------------------------------------------------
\begin{abstract}

Non-renormalisable versions of $SO(10)$\, based on irreducible representations with lesser degrees of freedom, are free of running into the catastrophe of non-perturbativity of standard model gauge couplings in contrast to the renormalisable versions having tensors with many degrees of freedom. $16_H$ is the smallest representation, participates in Yukawa Lagrangian at the non-renormalisable level, contributing to the charged and neutral fermion masses, and has six distinct scalars with different $B-L$ charges. We computed the leptoquark and diquark couplings of different pairs of scalars stemming from all possible decomposition of the term resulting from the coupling of $16_{H}$ with the ${\mathbf{16}}$ dimensional fermion multiplet of $SO(10)$,\, i.e. $\frac{{\mathbf{16}}\,{\mathbf{16}}\,16_{H}\,16_{H}}{\Lambda}$. Computing the tree and loop level contribution of different pairs to the effective dimension six, $B-L$ conserving operators, it turns out only three pairs, viz $\sigma\big(1,1,0\big)- T\big(3,1,\frac{1}{3}\big)$, and $H\big(1,2,-\frac{1}{2}\big)-\Delta\big(3,2,\frac{1}{6}\big)$, and $H-T$ can induce proton decay at tree level. Assuming that the Yukawa couplings of the $16_{H}$ are comparable to those of the $\overline{126}_{H}$ of a realistic $SO(10)$ model and setting the cutoff scale to the Planck scale typically constrains the $B-L$ breaking scale to be $4\sim 5$ orders of magnitude less than the cutoff scale $(\Lambda)$. Moreover, analysing the branching pattern of the leading two-body decay modes of the proton, we observed a preference for the proton to decay into second-generation mesons due to the hierarchical nature of Yukawa couplings. In a realistic $SO(10)$\, scenario, we find that $M_T >10^{8}$ TeV, while $M_\Delta$ could be as light as a few TeV$s$. 

\end{abstract}
%----------------------------------------------------------

\maketitle
%%%%%%%%%%%%%%%%%%%%%%%%%%%%%%%%%%%%%%%%%%%%%%%%%%%%%%%%%%%%

\section{Prelude}\label{sec:bg}
Even after the experimental verification of the Standard Model (SM) predictions with unprecedented accuracy, SM is still agnostic about the neutrino masses, baryon number violation, matter-antimatter asymmetry and many other observational and phenomenological pieces. Grand Unified Theory (GUT) is an aesthetically appealing and viable candidate that pragmatically solves some of the SM's inconsistencies while preserving its mores. GUT models place quarks and leptons within the same multiplet, which are otherwise scattered and independent in the SM. Various grand unification models based on different gauge symmetries have been known for more than four decades \cite{Georgi:1974sy,Fritzsch:1974nn,Pati:1973uk}.

Grand unified models based on \so\, gauge symmetry accommodates an entire generation of SM chiral fermions and a right-handed neutrino in a single sixteen-dimensional irreducible representation (irrep) ($\fs$-plet) (cf. \cite{Langacker:1980js, Aulakh:2002zr,Fukuyama:2004ps,Nath:2006ut} for review). The necessity of the presence of the right-handed neutrino is in alignment with the demand of light-neutrinos being massive, making the theory left-right symmetric above the so called seesaw scale. In addition to $\fs$-plet, \so\, GUT also comprises other scalar irreps pertinent to meet some elementary demands, i.e. 1) to decompose into the standard model, and 2) to reproduce the observed charged and neutral fermion mass spectrum and mixing angles. Three different kinds of irreps can, in principle, contribute to the \so\, scalar potential
i) \so\, gauge symmetry breaking irreps which are $45_{H},$ $54_{H}$ and $210_{H}$ \cite{Slansky:1981yr}, ii) irreps which induce rank reduction, i.e. $16_H,\, \overline{126}_H$, and are linked to neutrino mass generation \cite{Georgi:1979dq,Barbieri:1979ag,Buccella:1980qb,Yasue:1981nd}\footnote{Neutrino masses originating from the inclusion of $120_{H}$ instead of $16_{H}$ and $\overline{126}_{H}$ have been discussed in \cite{Barbieri:1979eh}.}, and iii) irreps participating in the Yukawa interactions at renormalisable level, $10_{H}$, $\overline{126}_{H}$ and $120_{H}$, and at least two of the aforementioned representations are needed for a realistic fermion mass spectrum \cite{Babu:1992ia,Bajc:2005zf,Joshipura:2011nn,Altarelli:2013aqa,Dueck:2013gca,Meloni:2014rga,Meloni:2016rnt,Babu:2016bmy,Ohlsson:2018qpt}.  Any GUT model utilizes at least one irreps from each case mentioned in (i, ii, and iii) for a realistic model, and the choice is rather subjective and influenced by various phenomenological aspects.

Incorporating the typically large tensor representations (LTR) ($120_{H}$, $\overline{126}_{H}$ and $210_{H}$) in the scalar sector accounts for the associated low-energy phenomenology while some of them contributing in the Yukawa potential at the renormalisable level. However, these large tensor representations contribute to the $\beta$-functions, causing low-energy gauge couplings to increase with energy. Consequently, the theory becomes strongly coupled in the ultraviolet (UV), potentially leading to the gauge couplings hitting the Landau pole before the Planck Scale ($M_{p}$). This phenomenon, known as asymptotic slavery (non-asymptotic freedom) in \so\, gauge theories, results in low-energy gauge couplings becoming unrelated to the non-perturbative evolution of gauge couplings in the UV, a feature that is uncalled for \cite{Maiani:1977cg,Vaughn:1978st,Rubakov:1999mu,Aulakh:2002ww,Aulakh:2002ph,Aulakh:2003kg,Bajc:2016efj}. In order to evade this concern, one can stick to smaller tensor representations ($10_{H}$, $16_{H}$, $45_{H}$ and $54_{H}$) and adhere to non-renormalisable interactions \cite{Babu:1994dc,Dvali:1996wh,Barr:1997hq,Albright:2002np,Chang:2004pb}. Non-renormalisable interactions could arise from the presence of another theory valid above a particular cutoff scale, of which the germane gauge theory could be a low-energy approximation. Likewise, the SM relies on renormalisable interactions to explain the masses of charged fermions and resorts to non-renormalisable interactions for the masses of neutral fermions, suggesting the existence of a more fundamental theory at higher energy scales. Moreover, incorporating higher-dimensional operators is also known to yield desired and viable Yukawa ratios among different generations of fermions \cite{Anderson:1993fe,Antusch:2019avd}, leading to a minimal model in terms of free parameters. Realistic $SO(10)$ models composed of smaller tensor representations have been considered in~\cite{Anastaze:1983zk,Preda:2022izo}.

The smallest representation capable of generating charged and neutral fermion masses, including the Majorana mass for the light and heavy neutrinos at the non-renormalisable level, is the $16_{H}$ representation. It comprises six distinct scalar particles, which are the scalar counterparts of the SM fermions residing in the $\fs$-dimensional fermion multiplet. Besides, these distinct particles are also non-trivially charged under $B-L$ symmetry and, thereby, are capable of mediating baryon (B) and lepton (L) number-changing phenomena like proton decay \cite{PhysRevLett.43.1566,PhysRevLett.43.1571,Sakai:1981pk,Golowich:1981sb,Nandi:1982ew,Dorsner:2004jj,Dorsner:2004xa,Kolesova:2016ibq,Buchmuller:2019ipg,Patel:2022wya}. The inclusion of only $16_{H}$ irrep along with the $\fs$-plet fermion, when decomposed, yields the interaction of two scalars with two SM fermions suppressed by a cutoff scale. In the absence of any LTR, these Yukawa couplings of $16_{H}$ can be of a similar order to those of an irrep coupling at the renormalisable level (preferably $\overline{126}_{H}$) to yield a realistic fermion mass spectrum and thus one can constrain the spectrum of the pair of scalars from various phenomenological implications with these determinable Yukawa couplings of $16_{H}$. 

This article is a continuation of our previous works in which we have constrained the spectrum of different scalars stemming from irreps capable of participating in the renormalisable $SO(10)$ Yukawa interactions. We have comprehensively classified scalars capable of inducing $B-L$ conserving and violating two body proton decays \cite{Patel:2022wya} and examined the role played by sextet scalars in different phenomenological implications \cite{Patel:2022nek}. This work considers the implications of scalars participating in non-renormalisable \so Yukawa interactions. We thoroughly analyse different pairs of scalars capable of inducing tree and loop-level two body proton decay emanating from all possible allowed decomposition of $\fs\,\fs\,16_{H}\,16_{H}$ coupling and thereby constraining various mediators from the same. Further, we determine the extent to which a pair of scalars could be light while adhering to the constraints imposed by proton decays in a typical realistic non-renormalisable \so\, model. A similar study has been performed in the context of supersymmetric non-renormalisable $SO(10)$ GUT in \cite{Wiesenfeldt:2005zx} and recently, proton decay, both at the loop level \cite{Helo:2019yqp,Gargalionis:2024nij} and via higher-dimensional operators \cite{Beneito:2023xbk}, has gained attention.

This article is organised as follows. In section (\ref{sec:couplings}), we explore the decomposition of the coupling between $16_{H}$ and $\fs$-plet fermions occurring at the non-renormalisable level. Additionally, we identify pairs of scalars within $16_{H}$ capable of inducing proton decay. The computation of the contribution from various scalar pairs to the effective proton decay operators is presented in section (\ref{sec:pdeff}). Utilizing these general results, we analyze the branching pattern of the leading modes of proton decay and constrain the mediators accordingly in section (\ref{sec:pds}). Finally, we conclude our study in section (\ref{sec:conc}).

\section{Couplings of $16_H$ in \so}{\label{sec:couplings}}
\label{sec:couplings}
A single $\mathbf{16}$-dimensional irreducible representation (irrep) of \so\, can accommodate an entire generation of standard model Weyl fermions along with their hypothetically-missing ally, a conjugated-right handed neutrino. Once the \so\, gauge symmetry is broken into one of its subgroups, i.e. $SU(5)\times U(1)$, the $16$-plet can be decomposed into the irreps of $SU(5)$ as shown below \cite{Gell-Mann:1979vob}.

\be {\label{eq:16inSU5}}
 \mathbf{16} = \big(\fn, \mathbf{-5}\big) + \big(\ff,\mathbf{3}\big) + \big(\ft,\mathbf{-1}\big)
\ee

The $SU(5)$ irreps shown in eq. (\ref{eq:16inSU5}) are the matter multiplets and consist of SM quarks and leptons. The SM Weyl fermions residing in the different $SU(5)$ irreps are shown below. Throughout this article, we denote generation indices by the uppercase Latin alphabets $\big(1\leq A,B,C...\leq3\big)$, $SU(5)$ indices by lowercase Latin letters $ \big( 1 \leq n , o , p,...,z   \leq 5 \big) $,  $SU(3)$ indices by lowercase Greek letters $\big( 1 \leq \alpha, \beta, \gamma... \leq 3 \big)$, and lowercase Latin alphabets $\big( 4\leq a, b, c,...,m \leq 5)$ represent the $SU(2)$ indices, unless specified explicitly. Moreover, the bold letter denotes the fermion representation, while the scalar representations are shown with subscript $H$.  

\beqa {\label{eq:SU(5)toSMfermions}}
\fn&=& \nu^C,\hspace{1cm} \ff_a\;=\; \varepsilon^{ab}\,l^{b},\hspace{1cm} \ff_{\alpha}\;=\;d^C_{\alpha},\nl 
\ft^{ab}&=& \varepsilon^{ab}\,e^C,\hspace{1cm} \ft^{a\,\alpha}\;=\;q^{a\alpha},\hspace{1cm}\ft^{\alpha\beta}\;=\;\varepsilon^{\alpha\beta\gamma}\,u^C_{\gamma}.
\eeqa

The two rank $SU(2)\;Levi-Civita$ tensor used in eq. (\ref{eq:SU(5)toSMfermions}) is defined as $\varepsilon_{45}=-\varepsilon_{54}=-\varepsilon^{45}=\varepsilon^{54}=1$, while the three rank $SU(3)\;Levi-Civita$ tensor is defined as $\varepsilon_{123}=\varepsilon^{123}=1$, and similarly for all other cyclic permutations. Additionally, $C$ is the charge conjugation matrix in the Lorentz space and is defined as $C \equiv i\sigma_2$, $\sigma_2$ being the Pauli matrix.

Similarly, the canonically normalised decomposition of scalars residing in the $16_H$-dimensional scalar representation is as follows \cite{Patel:2022wya}.

\beqa {\label{eq:SU(5)toSMscalars}}
1_H&=& \sigma,\hspace{1cm} 5^{\dagger}_{a\,\,H}\;=\; H^{\dagger}_a,\hspace{1cm} 5^{\dagger}_{\alpha\,\,H}\;=\;T^{\dagger}_{\alpha},\nl 
10^{ab}_H&=& t^{ab},\,\hspace{1cm} 10^{a\,\alpha}\;=\; \frac{1}{\sqrt{2}}\,\Delta^{a\alpha},\hspace{1cm}10^{\alpha\beta}\;=\;\Theta^{\alpha\beta}.
\eeqa

$16_H$ is the smallest representation, contributing to charged fermion masses and neutrino masses; however, it contributes to the Yukawa Lagrangian at the non-renormalisable level. Table (\ref{tab:tab1}) depicts the different scalar contained in $16_{H}$, together with their SM charge and $B-L$ quantum number. The $B-L$ of any multiplet is related to subgroups \, $U_1^X$, and $U_1^Y$ contained in \so\, by the following relation \cite{Buchmuller:2019ipg}.

\be {\label{eq:B-Ldef}}
B-L = \frac{4}{5}\,Y\,-\,\frac{1}{5}X.
\ee
where, $Y$ and $X$ are respectively the hypercharge and $U_1^X$ charge of the multiplet.

%%%%%%%%%%%%%%%%%%%%%%%%%%%%%%%%%%%%%%%%%%%%%%%%%%%%%
\begin{table}[t]
    \centering

    \begin{tabular}{ccc}
    \hline
    ~~~~SM Charges~~~~     & ~~~~Notation~~~~ & ~~~~$B-L$ Charges~~~~ \\
    \hline\hline
    $\left(1,1,0\right)$     & $\sigma$ & $1$ \\
    $\left(\overline{3},1,\sfrac{1}{3}\right)$     & $T^{\dagger}_{\alpha}$ & $\sfrac{-1}{3}$ \\
    $\left(1,2,\sfrac{-1}{2}\right)$     & $H^{\dagger}_a$ &$ -1$\\
    $\left(3,2,\sfrac{1}{6}\right)$     & $\Delta^{a\alpha}$ &  $\sfrac{1}{3}$\\
    $\left(\overline{3},1,\sfrac{-2}{3}\right)$     & $\Theta^{\alpha\beta}$ & $\sfrac{-1}{3}$ \\
    $\left(1,1,-1\right)$ & $t^{ab}$   & $1$  \\
    \hline

    \end{tabular}

    \caption{Classification of scalars present in $16_H$ alongwith their SM and $B-L$ charges}
    \label{tab:tab1}
\end{table}
%%%%%%%%%%%%%%%%%%%%%%%%%%%%%%%%%%%%%%%%%%%%%%%%%%%%%%%%%%%%%%%%%%%%%
We begin with an \so\, lagrangian with $\fs$ dimensional fermion multiplet and $16_{H}$ scalar representation. The Yukawa Lagrangian of $\mathbf{16}$ with $16_H$ to the first order in $\Lambda$ is shown below, where $\Lambda$ being the typical cutoff scale. 

\beqa {\label{eq:NRSO10}}
{\mathcal{L}} &\supset&  \frac{1}{\Lambda} y_{AB}\,\big(\mathbf{16}_{A}\,\mathbf{16}_{B}\;16_{H}\,16_{H}\big)\,+\, \frac{1}{\Lambda} \bar{y}_{AB}\,\big(\mathbf{16}_{A}\,\mathbf{16}_{B}\;16^{\dagger}_{H}\,16^{\dagger}_{H}\big) \,\hc
\eeqa

To decompose the Lagrangian given in eq. (\ref{eq:NRSO10}), one considers all the effectively generated Yukawa couplings allowed by the integration of all possible irreps (scalar or fermions) consistent with the assumed gauge symmetry. The transformation properties of $16$ dimensional irrep under \so\, is given as given as follows \cite{Slansky:1981yr}.

\beqa{\label{eq:16transform}}
16 \times 16 &=& 10 + 120 + 126^{\dagger}\nl
16 \times 16^{\dagger} &=& 1 + 45 + 210\eeqa
Eq. (\ref{eq:16transform}) suggests that once we have a term like the one in eq. (\ref{eq:NRSO10}), only a few possibilities of intermediate \so\, irreps are allowed by \so\, invariance. The exhaustive ways to generate the coupling given in eq. (\ref{eq:NRSO10}) is given as follows. In the following, we represent the coupling as $(...)_{D_1}(...)_{D_2}$ in a way that the representations inside the first (second) bracket are contracted to form the $D_1\, (D_2)$, such that $D_1 \times D_2 \supset \mathbf{1}$ where $\mathbf{1}$ transforms trivially under \so. Furthermore, $D_1\,(D_2)$ can be either a scalar or a fermion representation. 

\beqa {\label{eq:16Fs16Hs}}
      &\bullet &  \frac{h_{AB}}{\Lambda}\big(\fs_A\,\fs_B\big)_{10_H}\,\big(16_H\,16_H\big)_{10_H},\; \frac{\bar{h}_{AB}}{\Lambda}\big(\fs_A\,\fs_B\big)_{10_H}\,\big(16^{\dagger}_H\,16^{\dagger}_H\big)_{10^{\dagger}_H}, \nl  
     & & \frac{\tilde{h}_{AB}}{\Lambda}\big(\fs_A\,16_H\big)_{\mathbf{10}}\,\big(\fs_B\,16_H\big)_{\mathbf{10}}, \text{and}\;\frac{\hat{h}_{AB}}{\Lambda}\Big(\fs_{A}\,16_H\big)_{\ft}\,\Big(\fs_{B}\,16_H\Big)^{*}_{\ft^{\dagger}}\nl 
      & \bullet &  \frac{g_{AB}}{\Lambda}\big(\fs_A\,\fs_B\big)_{120_H}\,\big(16_H\,16_H\big)_{120_H},\; \frac{\bar{g}_{AB}}{\Lambda}\big(\fs_A\,\fs_B\big)_{120_H}\,\big(16^{\dagger}_H\,16^{\dagger}_H\big)_{120^{\dagger}_H}, \nl
     & &  \frac{\tilde{g}_{AB}}{\Lambda}\big(\fs_A\,16_H\big)_{\mathbf{120}}\,\big(\fs_B\,16_H\big)_{\mathbf{120}} \;\text{and}\; \frac{\hat{g}_{AB}}{\Lambda}\big(\fs_A\,16_H\big)_{\mathbf{120}}\,\big(\fs_B\,16_H\big)^*_{\mathbf{120^{\dagger}}} \nl
     &\bullet &   \frac{\bar{f}_{AB}}{\Lambda}\big(\fs_A\,\,\fs_B\big)_{126^{\dagger}_H}\,\big(16^{\dagger}_H\,16^{\dagger}_H\big)_{126_H}\;\text{and} \;\frac{\hat{f}_{AB}}{\Lambda}\big(\fs_A\,\,16_H\big)_{\mathbf{126}^{\dagger}}\,\big(\fs_B\,16_H\big)^*_{\mathbf{126}} \nl
     & \bullet & \frac{k_{AB}}{\Lambda}\, \big(\fs_A\,16^{\dagger}_H\big)_{\mathbf{1}}\,\big(\fs_B\,16^{\dagger}_H\big)_{\mathbf{1}},\;\frac{\tilde{k}_{AB}}{\Lambda}\, \big(\fs_A\,16^{\dagger}_H\big)_{\mathbf{45}}\,\big(\fs_B\,16^{\dagger}_H\big)_{\mathbf{45}},\nl 
     & & \text{and}\; \frac{\bar{k}_{AB}}{\Lambda}\, \big(\fs_A\,16^{\dagger}_H\big)_{\mathbf{210}}\,\big(\fs_B\,16^{\dagger}_H\big)_{\mathbf{210}}
\eeqa    
where, $ h,\, \bar{h}, \tilde{h}, \,\hat{h}, \,g, \, \bar{g}, \tilde{g},\, \hat{g},\,\bar{f}, \, \hat{f},\,k,\,\tilde{k}$ and $\bar{k}$ are various Yukawa couplings and $\Lambda$ corresponds to the energy scale where the intermediate \so\, irrep has been integrated out. It is also evident that less cases are possible when the intermediate integrated out irrep is $\overline{126}$ dimensional, as $\overline{126}\times \overline{126}$ does not contain $SO(10)$ singlet. It is to be noted that in Eq.~\eqref{eq:16Fs16Hs}, we have not considered the combination $\big(\fs-\fs^{\dagger}\big)_{1,\,45,\,210}\,$\,$\big(...\big)_{1,\,45,\,210}$ as the coupling $\big(\fs-\fs^{\dagger}\big)$ is conventionally a part of the gauge sector. We shall constrain ourselves to those possibilities which contribute to the Yukawa sector.

Having singled out all possible ways for generating the effective term mentioned in eq. (\ref{eq:NRSO10}), we further shift to decompose these \so\, invariant couplings, as given in eq. (\ref{eq:16transform}), into the SU(5) framework, starting with the parameterisation of the coupling between the $\fs$-plet fermion and the $10_{H}$ \cite{Nath:2001uw}.
 \beqa{\label{eq:SO10/10}}
 -{\cal{L}}^{10}_{Y} &\supset& H_{AB}\,\fs_{A}^T\,\cc\,\fs_{B}\,10_H \nl
&=&  i 2\sqrt{2}\, H_{AB}\,\left({\ft^{pq}_A}^T\, \cc\,\ff_{p\,B}\, \tilde{5}^{\dagger}_{q\,H}\, +\, \frac{1}{8} \varepsilon_{pqrst}\, {\ft^{pq}}^T_A\, \cc\,\ft^{rs}_B\, \tilde{5}^t_{H}\, -\, \fn_A^T\, \cc\,\ff_{p\,B}\, \tilde{5}^p_{H} \right)\nl &+& \rm{H.c.}
 \eeqa
 where, the $H_{AB}$ is constrained by its symmetricity in the generation space. In addition, the scalar fields belonging to the $10_H$ representation are denoted by a tilde placed over them. Furthermore, the decomposition of the interaction between the $16_{H}$ and $10_{H}$ representations can be expressed analogously to eq. (\ref{eq:SO10/10}), as illustrated below.

\beqa{\label{eq:SO10/10H}} 
 -{\cal{L}}^{10}_{Y} &\supset&  \eta\,16_{H}\,16_{H}\,10_H + M^2_{10_{H}}\,10^{\dagger}_{H}\,10_{H}\, \nl
&=&  i 2\sqrt{2}\, \eta\,\left(10_{H}^{pq}\,5^{\dagger}_{p\,H}\, \tilde{5}^{\dagger}_{q\,H}\, +\, \frac{1}{8} \varepsilon_{pqrst}\, {10_H^{pq}}\, 5_H^{rs}\, \tilde{5}^t_{H}\, -\, 1_H\,5^{\dagger}_{p}\, \tilde{5}^p_{H} \right)\nl
&+& M^2_{10_{H}}\,\tilde{5}^{\dagger}_H\,\tilde{5}_H \hc
\eeqa

To integrate out the $10_H$ representation, we utilise their couplings derived in eqs. (\ref{eq:SO10/10}) and (\ref{eq:SO10/10H}) in their irreducible $SU(5)$ components.  By utilizing equations of motion, we solve for the various $SU(5)$ components of $10_H$ in eq. (\ref{eq:SO10/10H}) and subsequently substitute the results into eq. (\ref{eq:SO10/10}) \cite{Nath:2001yj}. Additionally, it is imperative to mention that we integrate out representations with the same $U_1^X$ charge. For instance,  the $U_1^X$ charge of $\tilde{5}^{\dagger}_H \subset 16_{H}$ is $3$, while the $U_1^X$ charge of $5^{\dagger}_H \subset 10_{H}$ is $-2$ \cite{Slansky:1981yr,Feger:2019tvk}. Thus, identical $SU(5)$ representations belonging to different $SO(10)$ representations can differ in their $U_1^X$ charges. Consequently, we denote the $SU(5)$ scalars (and the fermions) intended for integration by placing a tilde ($\sim$) over them.  

\beqa{\label{eq:h10}}
\lag_{\rm{NR}} & \supset& \frac{h_{AB}}{\Lambda} \Big( \fs^T_{A}\,\cc\, \fs_B\Big)_{10}\,\Big(16_H\,16_{H}\Big)_{10} \nl
&=& \frac{-8\,h_{AB}}{\Lambda}\Big[ \big(\ft^{pq\,T}_A\,\cc\, \ff_{p\,B} \big)\,\big(-1_H\, 5^{\dagger}_{q\,H} + \frac{1}{8} \varepsilon_{qrstu}\,10^{rs}_H\,10^{tu}_H\big) + \big(-\fn^T_A\,\cc\,\ff_{p\,B} \nl &+& \frac{1}{8}\,\varepsilon_{qrstp}\,\ft^{qr\,T}_A\,\cc\ft^{st}_B \big)\,\big(10^{pw}_H\,5^{\dagger}_{w\,H}\big)  \Big] \hc
\eeqa
where, we set $\frac{\eta\,H_{AB}}{M^2_{10_{H}}}\,\,\equiv\,\frac{h_{AB}}{\Lambda}$. Similarly, we can derive the decomposition of term associated with $\bar{h}$ coupling of eq. (\ref{eq:16Fs16Hs}) as shown below.

\beqa{\label{eq:h10b}}
-\lag_{\rm{NR}} &\supset & \frac{\bar{h}_{AB}}                 {\Lambda}\Big( \fs^T_{A}\,\cc\, \fs_B\Big)_{10}\,\Big(16^{\dagger}_H\,16^{\dagger}_{H}\Big)_{10^{\dagger}}\nl
&=& \frac{8\,\bar{h}_{AB}}{\Lambda}\Big[ \big(\ft^{pq\,T}_A\,\cc\, \ff_{p\,B} \big)\,\big(10^{\dagger}_{qw\,H}\,5^{w}_H\big) \nl 
          &+& \big(-\fn^T_A\,\cc\,\ff_{p\,B} + \frac{1}{8}\,\varepsilon_{qrstp}\,\ft^{qr\,T}_A\,\cc\ft^{st}_B \big)\,\big(-1_H\, 5^{p}_H + \frac{1}{8} \varepsilon^{prstu}\,10^{\dagger}_{rs\,H}\,10^{\dagger}_{tu\,H}\big)  \Big]\nl
          &+& \rm{H.c.}
\eeqa

Further, the decomposition of $\fs$-plet fermion with $\ft$ dimensional fermion is shown below \cite{Nath:2001uw}.
\beqa{\label{eq:h10F}}
\lag_{Y}^{10} &\supset & \tilde{H}_{AB}\, \fs^T_{A}\,\cc\,\ft_B\, 16_H\,\nl
&=& i 2\sqrt{2}\, \tilde{H}_{AB}\,\Big({\ft^{pq}_A}^T\, \cc\,\tilde{\ff}_{p\,B}\, 5^{\dagger}_{q\,H}\, +\, \frac{1}{8} \varepsilon_{pqrst}\, {\ft^{pq}}^T_A\, \cc\,\big(\tilde{\ff}\big)^{*\,r}_{B}\, 10^{st}_{H}\, \nl
&-&\, \fn_A^T\, \cc\,\big(\tilde{\ff}\big)^{*\,p}_{B}\, 5^{\dagger}_{p\,H} \Big) \hc
\eeqa
where, the fermion representations belonging to $\ft$ are shown by putting tilde ($\sim$) over them. Utilizing eq. (\ref{eq:h10F}), the estimation of higher-mass dimensional terms with Yukawa coefficients $\tilde{h}$ and $\hat{h}$, mentioned in eq. (\ref{eq:16Fs16Hs}), is provided as follows:

\beqa{\label{eq:hAB}}
\lag_{NR} &\supset &  \frac{\tilde{h}_{AB}}{\Lambda}\Big( \fs^T_{A}\,\cc\, 16_H\Big)_{\ft}\,\Big( \fs^T_{B}\,\cc\, 16_H\Big)_{\ft} \nl &+& \frac{\hat{h}_{AB}}{\Lambda}\Big(\fs^T_{A}\, \cc 16_H\big)_{\ft}\,\Big(\fs^T_{B}\,\cc\,16_H\Big)^{*}_{\ft^{\dagger}}  \hc\nl  
          &=& 
          \Big[\frac{-8\,\tilde{h}_{AB}}{\Lambda} \big(\ft^{pq\,T}_A\,\cc\, 5^{\dagger}_{p\,H} \big)\,\big(-\fn_B\, 5^{\dagger}_{q\,H} + \frac{1}{8} \varepsilon_{qrstu}\,\ft_B^{rs}\,10^{tu}_H\big)\,+ A\leftrightarrow B \Big]\,\nl
          &+& \frac{\hat{h}_{AB}}{\Lambda}\Big[ \big(\ft^{pq\,T}_{A}\cc\,5_{r\,H}^{\dagger}\big)\, \big(\ft^{pq}_{B}\,\,5_{r\,H}^{\dagger}\big)^* + \big(\ff^T_{p\,A}\cc\,10^{qr}_H\big)\, \big(\ff_{p B}\,10^{qr}_H\big)^* \nl
          &+& \big( -\fn^T_A \cc\, 5^{\dagger}_{t\,H} + \frac{1}{8}\varepsilon_{pqrst} \, \ft^{pq\,T}_A\cc 10^{rs}_H\big)\,\big( -\fn_B \, 5^{\dagger}_{t\,H} + \frac{1}{8}\varepsilon_{xywzt} \, \ft^{xy}_B 10^{wz}_H\big)^*\Big]\ \nl &+ & \rm{H.c.}
\eeqa

The expressions provided in eqs. (\ref{eq:h10}), (\ref{eq:h10b}), and (\ref{eq:hAB}) collectively represent the decomposition of higher-dimensional terms involving an intermediate integrated-out irrep of dimension $10$, which can be either scalar or fermionic. By using the embedding of SM fermions into various irreducible representations of $SU(5)$, as given in eq. (\ref{eq:SU(5)toSMfermions}), and the embedding of different scalars into different $SU(5)$ irreps of $16_{H}$ as described in eq. (\ref{eq:SU(5)toSMscalars}), these expressions can be further decomposed into couplings involving two SM fermions and two scalars, preserving the SM gauge symmetry.

We proceed to calculate the decomposition of effective terms permitted by the integration of $120$-dimensional irrep. Repeating the exercise conducted for the case of the $10$-dimensional irrep, we obtain the following expression.

\beqa {\label{eq:gAB}}
\lag_{NR} &\supset & \frac{g_{AB}}{\Lambda}\Big( \fs^T_{A}\,\cc\, \fs_B\Big)_{120}\,\Big(16_H\,16_{H}\Big)_{120}+ \frac{\bar{g}_{AB}}{\Lambda}            \Big( \fs^T_{A}\,\cc\, \fs_B\Big)_{120}\,\Big(16^{\dagger}_H\,16^{\dagger}_{H}\Big)_{120^{\dagger}}\,\nl   
            &+& \frac{\tilde{g}_{AB}}{\Lambda}\Big( \fs^T_{A}\,\cc\, 16_H\Big)_{\mathbf{120}}\,\Big( \fs_{B}\,\cc\, 16_H\Big)_{\mathbf{120}} \nl
            &+& \frac{\hat{g}_{AB}}{\Lambda}\Big(\fs^T_{A}\, \cc 16_H\big)_{\mathbf{120}}\,\Big(\fs_{B}\,\cc\,16_H\Big)^{*}_{\mathbf{120}^{\dagger}}
          \hc \nl 
          &=& \frac{-4\,g_{AB}}{3\Lambda}\,\Big[ 2 \big(\fn_A\,\cc\,\ff_{p\,B}\big)\,\big(10^{pq}_H\,5^{\dagger}_{q\,H}\big) - 4 \big(\ft_A^{pq\,T}\,\cc\,\ff_{p\,B}\big)\,\big(1_H\,5^{\dagger\,q}_H\big) \nl
          &+& 2 \big(\ff^T_{p\,A}\,\cc \ff_{q\,B}\big)\,\big(1_H\,10^{pq}_H\big)
          + 2 \big(\fn_A\,\cc\,10^{pq}_B\big)\,\big(5^{\dagger}_{p\,H}\,5^{\dagger}_{q\,H}\big)\nl
          &-&\frac{1}{2} \big(\varepsilon_{pqrst}\,\ft^{pq\,T}_{A}\,\cc\,\ft^{rn}_B\big)\,\big(5^{\dagger}_{n\,H}\,10^{st}_H\big)
          - \frac{1}{2} \big(\ff_{A\,n}\,\ft^{pq}_B\big)\,\big(\varepsilon_{pqrst}\,10^{rs}_H\,10^{tn}_H\big)\Big] \,\nl 
          &+& \frac{4\bar{g}_{AB}}{3\Lambda}\,\Big[ 4\big( \fn_{p\,A}^{T}\cc\,\ff_B\big)\,\big(1_H\,5^{p}_H\big)+ \big(\ft^{pq\,T}_A\cc\,\ff_{r\,B}\big)\,\big(10^{\dagger}_{pq\,H}\,5^{r}_H\big) \nl
          &+& 2 \big(\ff_{p\,A}\cc\,\ff_{q\,B}\big)\,\big(5^p_H\,5^q_H\big) + 2 \big(\fn_A^T\cc\,\ft^{pq}_{B}\big)\,\big(1_H\,10^{\dagger}_{pq\,H}\big) + 2 \big(\ff_{p\,A}\cc\,\ft_B^{qr}\big)\,\big(5^{p}_H\,10^{\dagger}_{qr}\big)\nl
          &+& \frac{1}{8} \big(\varepsilon_{pqrst}\ft_A^{pq\,T}\cc\,\ft_B^{rn}\big)\,\big(\varepsilon^{xyzst}10^{\dagger}_{xy\,H}\,10^{\dagger}_{zn\,H}\big)\Big]\nl
          &-& \frac{4\tilde{g}_{AB}}{3\,\Lambda}\Big[ -2\big( \fn_A^T\,\cc 5^{\dagger}_{p\,H}\big)\,\big(\ft^{pq}_B\,5^{\dagger}_{q\,H}\big)-2 \big( \ff^T_{p\,A}\cc\,5^{\dagger}_{q\,H}\big)\,\big(\fn_B\,10^{pq}_H\big) \nl
          &-& \frac{1}{2}\big(\varepsilon_{pqrst}\,\ft_A^{pq\,T}\,\cc\,10^{rn}_H\big)\,\big(\ff_{n\,B}\,10^{st}_H\big)\Big]\nl
          &+& \frac{4\hat{g}_{AB}}{3\,\Lambda}\Big[ \big(\ft^{pq\,T}_A\cc\,5^{\dagger}_{p\,H}\big)\,(\ft_B^{sq}\,5^{\dagger}_{s\,H}\big)^* + 2\big(\ff^T_{p\,A}\cc\,10^{qr}_{H}\big)\,\big(\ff_{p\,B}\,\,10^{qr}_{H}\big)^* \nl
          &+& \big(\ff^T_{p\,A}\cc\,5^{\dagger}_{q\,H}\big)\,\big(\ff_{p\,B}\,\,5^{\dagger}_{q\,H}\big)^* + \frac{1}{8}\,\big(\varepsilon_{pqrst}\,\ft^{pq\,T}_A\,\cc\,10^{rn}_H\big)\,\big(\varepsilon_{xyzst}\,\ft^{xy}_B\,\,\,10^{zn}_H\big)^*\Big] \nl & +& \rm{H.c.} 
\eeqa
Eq. (\ref{eq:gAB}) provides the coupling of two $SU(5)$ fermion multiplets and two scalar multiplets resulting from the integration of the $120$ representation. It recieves contributions from $g$, $\bar{g}$, $\tilde{g}$, and $\hat{g}$, where the former two couplings are antisymmetric in nature.

We now turn to the computation of terms when the intermediate integrated out irrep is $126$ dimensional. In contrast to the case of $10_{H}$ and $120_H$, fewer possibilities of decomposition of the term mentioned in eq. (\ref{eq:NRSO10}) mediated by $126$ are allowed. The latter is due to the transformation property of $126$-dimensional irrep \cite{Gell-Mann:1979vob}.

Analogous to the case of $10_H$ and $120_H$, we calculate the decomposition of higher-dimensional terms comprising two $\fs$ and two $16_{H}$ representations, mediated by the $126$-dimensional irrep. This analysis considers both possibilities of the $126$ being either a scalar or a fermion.

\beqa {\label{eq:fAB}}
\lag_{NR}     & \supset &  \frac{\bar{f}_{AB}}{\Lambda}\big(\fs^T_A\,\cc\,\fs_B\big)_{126^{\dagger}_H}\,\big(16^{\dagger}_H\,16^{\dagger}                               _H\big)_{126_H} \nl
              &+& \frac{\hat{f}_{AB}}{\Lambda}\big(\fs_A^T\,\cc\,16_H\big)_{\mathbf{126}^{\dagger}}\,\big(\fs_B\,     16_H\big)^*_{\mathbf{126}}  \hc\nl
             &=& \frac{-4\,\bar{f}_{AB}}{15\,\Lambda}\,\Big[ \big(\fn_A^T\,\cc\,\ft_B^{pq}\big)\,\big(1_H\,10^{\dagger}_{pq\,H}\big) - \big(\ff_{p\,A}\cc\,\ff_{q\,B}\big)\,\big(5^p_H\,5^q_H\big) \nl
             &+& \frac{3}{2} \big(\fn_A^T\,\cc\,\ff_{p\,B} +\frac{1}{24} \varepsilon_{qrstp}\,\ft^{qr}_A\,\cc\,\ft^{st}_B\big)\,\big(1_H\,5^p_H + \frac{1}{24} \varepsilon^{uvwxp} 10^{\dagger}_{uv\,H}\,10^{\dagger}_{wx\,H}\big)\nl
             &+& \big(\ft_A^{pq\,T}\,\cc\,\ff_{r\,B}\big)\,\big(10^{\dagger}_{pq\,H}\,5^k_H\big) -  \frac{1}{4\sqrt{3}}\,\big(\varepsilon_{pqnuv}\, \ft_{A}^{pq\,T}\,\cc\,\ft_B^{rs}\big)\,\big(\varepsilon^{xynuv}\,10^{\dagger}_{xy\,H}\,10^{\dagger}_{rs\,H}\big) \Big]\nl
             &+& \frac{4\hat{f}_{AB}}{15\,\Lambda}\,\Big[ \big(\ff^T_{p\,A}\,\cc\,5^{\dagger}_q\big)\,\big(\ff_{p\,B}\,\cc\,5^{\dagger}_q\big)^* + \big(\ft^{pq\,T}_A\cc\,5^{\dagger}_{p\,H}\big)\,(\ft_B^{sq}\,5^{\dagger}_{s\,H}\big)^* \nl
             &+& \big(\ff^T_{p\,A}\cc\,10^{qr}_{H}\big)\,\big(\ff_{p\,B}\,\,10^{qr}_{H}\big)^* + \frac{1}{288}\,\big( \varepsilon_{pqtuv}\,\ft^{pq\,T}_A\,\cc\,10_H^{rs}\big)\,\big( \varepsilon_{wxtuv}\,\ft^{wx}_B\,10_H^{rs}\big)^* \nl
             &+&  \frac{1}{288}\,\big( \varepsilon_{pqtuv}\,\ft^{rs\,T}_A\,\cc\,10_H^{pq}\big)\,\big( \varepsilon_{wxtuv}\,\ft^{rs}_B\,10_H^{wx}\big)^* \Big] \hc 
\eeqa

We have, so far, computed the couplings where two alike $16$-plets transform. Now, we consider the case when one $16$-plet couples to its conjugate, and the handful of resulting outcomes are possible and mentioned in eq. (\ref{eq:16transform}). The irreducible representations resulting from the decomposition of $16$ and its conjugate are called the adjoint representations, which are real representations. Unlike the cases of $10$, $120$, and $126$, only the fermionic representations of these adjoint irreps will generate the expression presented in eq. (\ref{eq:NRSO10}). The interaction of $\fs$-plet with $\mathbf{45}$ is parameterised below \cite{Nath:2001yj}.

\beqa{\label{eq:1645}}
-\lag^{45}_{Y} &\supset & \tilde{k}_{AB}\, \fs^T_A\,\cc\,\mathbf{45}_B\,16^{\dagger}_H \nl
&=& \frac{i}{\sqrt{2}}\,\tilde{k}_{AB}\, \Bigg[\sqrt{5}\Big( \frac{3}{5} \ff_{p\,A}^T\,\cc\,5^p_H + \frac{1}{10}\, \ft_A^{pq\,T}\,\cc\, 10^{\dagger}_{pq\,H} - \fn_A^T\,\cc\,1^{\dagger}_H\Big) \tilde{\fn}_{B} \nl
&+& \Big(-\ft_A^{st\,T}\,\cc\,1_H + \frac{1}{2} \varepsilon^{pqrst} \ff_{p\,A}^T\,\cc\,10^{\dagger}_{qr\,H}\Big) \big(\tilde{\ft}\big)^{*}_{st\,B}\nl
&+& \Big( -\fn^T\,\cc\,10^{\dagger}_{st\,H} + \frac{1}{2} \varepsilon_{pqrst}\, \ft^{pq\,T}_A\,\cc\,5^r_{H} \Big) \tilde{\ft}^{st}_{B} \nl
&+& 2 \Big( \ft^{pr\,T}_A\,\cc\,10^{\dagger}_{rq} - \ff^T_{q\,A}\,\cc\,5^{p}\big)\,\tilde{\mathbf{24}}^q_{p\,B}\Bigg] \hc
\eeqa

Using the decomposition given in eq. (\ref{eq:1645}), the effective coupling between two $\fs$ and $16^{\dagger}_H$ mediated by $\mathbf{45}$ is computed as follows

\beqa {\label{eq:ktab}}
\lag_{\rm{NR}} &\supset & -\frac{\tilde{k}_{AB}}{\Lambda}\, \big(\fs^T_A\,\cc\,16^{\dagger}_H\big)_{\mathbf{45}}\,\big(\fs_B\,16^{\dagger}_H\big)_{\mathbf{45}} \nl
     &=& -\frac{\tilde{k}_{AB}}{2\Lambda}\Big[5\,\big( -\fn_A^T\,\cc\,1_H + \frac{1}{10}\,\ft_A^{pq\,T}\,\cc\,10^{\dagger}_{pq\,H} + \frac{3}{5}\, \ff^T_{p\,A}\,\cc\,5^{p}_H\big)\,\big( -\fn_B\,\,1_H \nl 
     &+& \frac{1}{10}\,\ft_B^{st}\,\,\,10^{\dagger}_{st\,H} +\frac{3}{5} \,\ff_{t\,B}\,\,\,5^{t}_H\big) +
      4\big(\ft_A^{pr\,T}\,\cc10^{\dagger}_{rq\,H}\big)\,\big(\ft^{qn}_B\,10^{\dagger}_{np\,H}\big)\nl 
      & - & 4 \big(\ft_A^{pr\,T}\,\cc10^{\dagger}_{rq\,H}\big)\,\big(\ff_{p\,B}\,5^q_H\big) -4  \big(\ff^T_{p\,A}\,\cc\,5^q_H\big)\,\big(\ft_B^{pr}\,\cc10^{\dagger}_{rq\,H}\big)\nl &+& 4 \big(\ff^T_{p\,A}\,\cc\,5^q_H\big)\,\big(\ff_{q\,B}\,\cc\,5^p_H\big) \nl
      &+& \big(-\ft_A^{pq\,T}\,\cc\,1_H + \frac{1}{2}\,\varepsilon^{rstpq}\,\ff_{A\,r}^{T}\,\cc\,10^{\dagger}_{st\,H}\big)\,\big(-\fn_B\,10^{\dagger}_{pq\,H} + \frac{1}{2} \varepsilon_{uvwpq}\ft^{uv}\,5^w_H\big) \Big]  \nl
      &+& \rm{H.c.}
\eeqa

 Analogous to eq. (\ref{eq:ktab}), one can compute the coupling of $\fs$ and $16^{\dagger}_H$ with $\mathbf{1}$ and $\mathbf{210}$ and obtain the effectively generated couplings mediated by them which are shown below. 
 
\beqa{\label{eq:kAB}}
\lag_{NR} &\supset& \frac{k_{AB}}{\Lambda}\, \big(\fs^T_A\,\cc\,16^{\dagger}_H\big)_{\mathbf{1}}\,\big(\fs_B\,16^{\dagger}_H\big)_{\mathbf{1}} 
+   \frac{\bar{k}_{AB}}{\Lambda}\, \big(\fs^T_A\,\cc\,16^{\dagger}_H\big)_{\mathbf{210}}\,\big(\fs_B\,16^{\dagger}_H\big)_{\mathbf{210}} \hc\nl
     &=& -\frac{k_{AB}}{\Lambda}\,\Big[\big( \fn_A^T\,\cc\,1_H + \frac{1}{2}\,\ft_A^{pq\,T}\,\cc\,10^{\dagger}_{pq\,H} - \ff_{p\,A}\,\cc\,5^{p}_H\big)\, \big( \fn_A\,\,1_H \nl &-&
       \frac{1}{2}\,\ft_B^{pq}\,\,10^{\dagger}_{pq\,H} - \ff_{p\,B}\,\,\,5^{p}_H\big)\Big] \nl
      &-& \frac{2\bar{k}_{AB}}{3}\,\Big[\frac{3}{2}\big(\ff^T_{p\,A}\,\cc\,5^q_H+\frac{1}{3}\ft^{qr\,T}_A\,\cc\,10^{\dagger}_{rp\,H}\big)\,\big(\ff_{q\,B}\,5^p+\frac{1}{3}\ft^{pn}_B\,10^{\dagger}_{nq\,H}\big)\nl
      &+& \frac{5}{8}\big(\fn_A^T\,\cc\,1_H \frac{1}{10}\ft^{pq\,T}\,\cc\,10^{\dagger}_{pq\,H}+\frac{1}{5}\ff^T_{p\,A}\,\cc\,5^p_H\big)\,\big(\fn_B\,1_H +\frac{1}{10}\ft^{st}_B\,10^{\dagger}_{st\,H}+\frac{1}{5}\ff_{s\,B}\,5^s_H\big)\nl
      &+& \frac{3}{4}  \big(\ft_A^{pq\,T}\,\cc\,1_H + \frac{1}{6}\,\varepsilon^{rstpq}\,\ff_{A\,r}^{T}\,\cc\,10^{\dagger}_{st\,H}\big)\,\big(\fn_B\,10^{\dagger}_{pq\,H} + \frac{1}{6} \varepsilon_{uvwpq}\ft^{uv}\,5^w_H\big)\nl
      &+& \frac{1}{36}\,\big(\varepsilon_{pqrst}\,\ft^{in\,T}_{A}\,\cc\,5^q_H\big)\,\big(\varepsilon^{uvrst}\,\ff_{u\,B}\,10^{\dagger}_{vn\,H}\big) +\frac{1}{16}\big(\ft^{pq\,T}_A\,\cc\,10^{\dagger}_{st\,H}\big)\,\big(\ft^{st}_B\,10^{\dagger}_{pq\,H}\big)\nl
      &+& \big(\ff_A^T\cc\,1_H\big)\,\big(\fn_B\,5^{i}_H\big) \Big] \hc
\eeqa

The $SU(5)$ invariant expressions given in eqs. (\ref{eq:h10}, \ref{eq:h10b}, \ref{eq:hAB}, \ref{eq:gAB}, \ref{eq:fAB}, \ref{eq:ktab}, and \ref{eq:kAB}) depict the different ways of arriving at the term given in eq. (\ref{eq:NRSO10}). It demonstrates that incorporating the $SO(10)$ invariant combination of two $\fs$-plets and $16_{H}$, along with its conjugate partner, at the leading non-renormalisable level, can be decomposed into several $SU(5)$ invariant expressions. The latter arises from the various possibilities of contracting the same $SO(10)$ invariant term. When further decomposed into expressions respecting the SM gauge symmetry, they yield the couplings of a pair of scalars from Table(\ref{tab:tab1}) with the SM fermions. The resulting dimension-5 coupling can have various phenomenological implications, ranging from contributing to fermion masses (charged and neutral) to baryon number-violating processes (nucleon decay).

We intend to constrain the spectrum of scalars residing in $16_{H}$ from nucleon decay. The respective scalar must have leptoquark and diquark couplings to mediate proton decay. In Table (\ref{tab:scalarclassification}), we gather the couplings of respective pairs of scalars with SM fermions obtained after the decomposition of eqs. (\ref{eq:h10}, \ref{eq:h10b}, \ref{eq:hAB}, \ref{eq:gAB}, \ref{eq:fAB}, \ref{eq:ktab}, and \ref{eq:kAB}). Furthermore, we are interested in pairs of scalars with both diquark and leptoquark vertices. Therefore, we only consider the coupling of scalar pairs capable of inducing proton decay, while suppressing the $SU(3)$, $SU(2)$, family indices. 

%%%%%%%%%%%%%%%%%%%%%%%%%%%%%%%%%%%%%%%%
\begin{table}[t]
    \centering
    \begin{tabular}{ccc}
    \hline
        ~~~Pair Of Scalar~~~ & ~~~Diquark~~~ & ~~~Leptoquark~~~ \\
    \hline\hline
     $\sigma-\sigma$     & \xmark & \xmark \\
     $\sigma-T$ & $q\,q; u^C\,d^C$  & $q\,l; u^C\,e^C$  \\
     $\sigma- H$& $q\,u^C;\,q\,d^C$ & \xmark \\
     $\sigma -\Delta$& \xmark & $d^C\,l$ \\
     $\sigma- \Theta$ & $d^C\,d^C$  & \xmark \\
     $\sigma-t$& \xmark & \xmark \\
     $T-T$& $d^C\,d^C$ & \xmark \\
     $T-H$& $q\,u^{C\,*}$ & $q\,e^{C\,*}$  \\
     $T-\Delta$& $q\,d^C$ & \xmark \\
     $T-\Theta$& $q\,q; u^C\,d^C$  & $q\,l; u^C\,e^C$\\
     $T-t$& \xmark & $d^C\,e^C$  \\
     $H-H$& $q\,q$;\, $d^C\,d^C$ & \xmark \\
     $H-\Delta$& $q\,q; u^C\,d^C$  & $q\,l; u^C\,e^C$\\
     $H-\Theta$& \xmark & $l\,u^C$ \\
     $H-t$& $q\,d^C$ & \xmark \\
     $\Delta-\Delta$& $q\,q;\;u^C\,d^C$ & $q\,l;\;u^C\,e^C$ \\
     $\Delta-\Theta$& $q\,u^{C\,*}$ & $q\,e^{C\,*}$ \\
     $\Delta-t$& $q\,u^{C\,*}$ & $q\,e^{C\,*}$   \\
     $\Theta-\Theta$& $u^C\,d^C$;\, $d^C\,d^C$ & \xmark  \\
     $\Theta-t$& $q\,q; u^C\,d^C$  & $q\,l; u^C\,e^C$\\
     $t-t$& \xmark & \xmark \\
     \hline
    \end{tabular}
    \caption{Coupling of different pair of scalars with the SM fermions. The ``\xmark"\, represent the respective coupling (diquark or leptoquark) is absent.}
    \label{tab:scalarclassification}
\end{table}

%%%%%%%%%%%%%%%%%%%%%%%%%%%%%%%%%%%%%%%%%%%%%%%%%%%%%%%%%%%%%%%%

A few comments are in order regarding table (\ref{tab:scalarclassification}).
\begin{itemize}
    \item The depiction of a pair of scalars in the table \ref{tab:scalarclassification} stands for all possible ways the scalars can appear in the eqs. (\ref{eq:h10}, \ref{eq:h10b}, \ref{eq:hAB}, \ref{eq:gAB}, \ref{eq:fAB}, \ref{eq:ktab}, and \ref{eq:kAB}), so $\sigma-T$ stands for $\sigma-T$, $\sigma-T^*$, $\sigma^*-T$ and $\sigma^*-T^*$.
    \item A scalar coupling with its conjugate partner, in principle, couples with any SM fermions with their conjugate partner; hence, those couplings are not mentioned in the table (\ref{tab:scalarclassification}).
  \item   The couplings of $\nu^C$ in the table (\ref{tab:scalarclassification}) have been omitted as they cannot contribute to proton decay at dimension 8.
  \item Out of twenty different pairs of scalars, only eight have the leptoquark and diquark coupling, hence can induce proton decay. These different pairs of scalars are $\sigma-T$, $T-H$, $T-\Delta$, $H-\Delta$, $\Delta-\Delta$, $\Delta-\Theta$, $\Delta-t$ and $\Theta-t$. 
   \item  It is apparent that $\Delta$ can couple to any scalar mentioned in Table (\ref{tab:tab1}), and can yield a diquark and leptoquark vertices.
\end{itemize}

After summarising the diquark and leptoquark vertices of different pairs of scalars resulting from the decomposition of various computed $SU(5)$ invariant expressions, we further focus on their explicit contributions to the various effective nucleon decay operators in terms of their Wilson coefficients.

\section{Dimension Six Effective Operators} {\label{sec:pdeff}}

The standard model gauge symmetry preserving and $B-L$ conserving (or $B+L$ violating) effective mass dimension-six ($D=6$) operators capable of inducing nucleon (free proton or a bound neutron) decay are shown below \cite{PhysRevLett.43.1566,PhysRevLett.43.1571,Sakai:1981pk}. As they conserve $B-L$, these operators are responsible for two-body nucleon decays, in which the nucleon decays into either charged anti-leptons or neutral ones, accompanied by mesons. The vertices leading to such operators are forbidden in the SM due to the apparent conservation of baryon and lepton numbers. Therefore, they result from the integration of degrees of freedom of particles (scalar or vector) which are not part of the SM.
\beqa \label{eq:bnc}
\mathcal{O}_{1} &=& c_1[A,B,C,D]\,\varepsilon_{\alpha\beta\gamma}\,\varepsilon_{gh}\,\varepsilon_{ab}\,\Big(q^{g\alpha\,T}_{A}\,\cc q_B^{h\,\beta}\Big)\,\Big( q^{a\,\gamma\,T}_C\,\cc l^b_D\Big),\nonumber\\
\mathcal{O}_2 &=& c_2[A,B,C,D]\,\varepsilon_{\alpha\beta\gamma}\,\varepsilon_{gh}\,\Big(q^{g\alpha\,T}_{A}\,\cc\, q_B^{h\,\beta}\Big)\,\Big( u^{C\,T}_{\,\gamma\,C}\,\cc e^{C}_D\Big)^{*},\nonumber\\
\mathcal{O}_3 &=& c_3[A,B,C,D]\,\varepsilon^{\alpha\beta\gamma}\,\varepsilon^{ab}\, \Big(u^{C\,T}_{\alpha\,A}\cc\, d^C_{\beta\,B}\Big)\,\Big(q^{a\,\gamma\,T}_C\cc\,l^b_D\Big)^*,\nonumber\\
\mathcal{O}_4 &=& c_4[A,B,C,D]\,\varepsilon^{\alpha\beta\gamma}\, \Big(u^{C\,T}_{\alpha\,A}\cc\, d^C_{\beta\,B}\Big)\,\Big( u^{C\,T}_{\,\gamma\,C}\,\cc e^{C}_D\Big).
\eeqa
The coefficients $c_{1,2,3,4}$ represent various Wilson coefficients associated with the newly introduced degrees of freedom and are the effective strength of the respective operator. Typically, these coefficients depend on the Yukawa couplings and are suppressed by the mass of the mediator. Additionally, $A$, $B$, $C$, and $D$ represent the generation indices of SM fermions involved in the operator. Only specific choices of $A$, $B$, $C$, and $D$ are permitted for tree-level nucleon decays. Specifically, only up-quarks can participate in the operator and for down-quarks and charged leptons, only fermions up to the second generation are allowed, while all three generations are allowed for neutral leptons. 

\begin{figure}[t]
    \centering
    \includegraphics[width=0.75\linewidth]{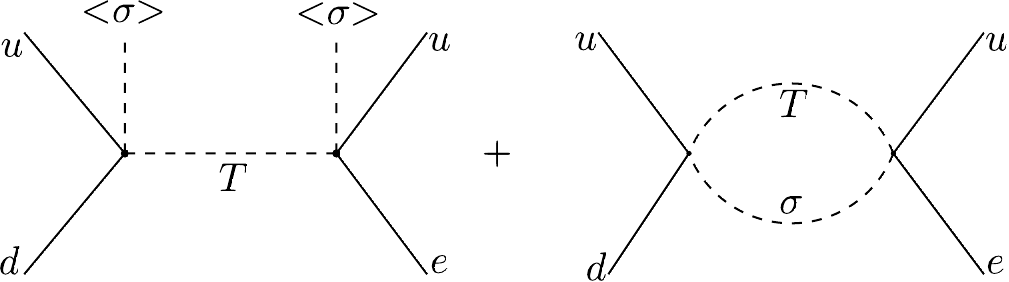}
    \caption{Tree and loop level proton decay topologies generated by the pair $\sigma-T$, drawn using \cite{Harlander:2020cyh}.}
    \label{fig:pdecayconfig}
\end{figure}

In the present scenario, the Lagrangian contains mass dimension-five terms that can generate the $D=6$ operators, mentioned in eq. (\ref{eq:bnc}), through two ways: 1) Integrating out a pair of scalars resulting in $D=8$ operators, with another pair of scalars simultaneously acquiring a vacuum expectation value (vev), resulting in an effective $D=6$ operator (cf. left diagram of Fig. (\ref{fig:pdecayconfig})); 2) The two pairs of scalars effectively generating  $D=6$ operators via loop mediation (cf. right diagram of Fig. (\ref{fig:pdecayconfig})). These $D=6$ effective operators can arise from specific terms of eqs. (\ref{eq:h10}, \ref{eq:h10b}, \ref{eq:hAB}, \ref{eq:gAB}, \ref{eq:fAB}, \ref{eq:ktab}, and \ref{eq:kAB}), or they can result from a linear combination of various terms. Integrating various scalars does not necessarily yield all the four operators mentioned in eq. (\ref{eq:bnc}). Moreover, there exists the possibility of other $B-L$ conserving effective operators at $D=8$, in addition to the operators mentioned in eq.(\ref{eq:bnc}) that can lead to proton decays involving multimeson modes \cite{Helo:2019yqp,Heeck:2019kgr}. These decays are inherently phase space suppressed compared to the two-body decays, and hence, are not explored in this analysis.

Below, we evaluate the contribution to the operators given in Eq. (\ref{eq:bnc}) mediated by various pairs of scalars. While deriving the contributions to the operators arising from different pairs of scalars, we have assumed only one particular \so\, irrep can mediate the decomposition of term given in eq. (\ref{eq:NRSO10}) at a time. However, in principle, similar operators arising from the same pair of scalars from the decomposition of different $\text{SO}(10)$-mediated irreps can mix, a possibility we remain ignorant of. 

%%%%%%%%%%%%%%%%%%%%%%%%%%%%%%%%%%%%%%%%%%%%%%%%%%%%%%%%%%%%

%%%%%%%%%%%%%%%%%%%%%%%%%%%%%%%%%%%%%%%%%%%%%%%%%%%%%%%%%%%%

We focus on the contribution of the $\sigma-T$ pair, capable of inducing proton decay by coupling to both leptoquark and diquark (cf. Table (\ref{tab:scalarclassification})). This pair induces proton decay at both tree and loop levels ( cf. Fig. (\ref{fig:pdecayconfig})). Integrating out the degrees of freedom of $T$ results in a $D=8$ operator. As $\sigma$ is a singlet under SM gauge symmetry, its vev reduces the $D=8$ effective operator to one of the $D=6$ operators in eq. (\ref{eq:bnc}). Moreover, this pair also mediates proton decay at the loop level with $\sigma$ and $T$ as scalars propagating inside the loop. The contribution to the operators, $\mathcal{O}_{1,2,3,4}$, in flavour basis mediated by $\sigma-T$ is as follows.

\beqa \label{eq:p1}
c_1\big[A,B,C,D\big] &=& \bigg(\frac{\langle\sigma\rangle ^2}{\Lambda^2\,M_T^2} + \frac{D\big(M^2_T,M^2_\sigma\big)}{\Lambda^2}\bigg)\,                           \Big[32 \bar{h}_{AB}\,h_{CD}   \Big],\nonumber\\
c_2\big[A,B,C,D\big] &=& \bigg(\frac{\langle\sigma\rangle^2}{\Lambda^2\,M_T^2} + \frac{D\big(M^2_T,M^2_\sigma\big)}{\Lambda^2}\bigg)\,                           \Big[ 32 \, \bar{h}_{AB}\, \bar{h}^*_{CD} - \frac{8}{2025} f_{AB}\, f^*_{CD}                                   \nonumber\\
                     &+&  
                      2\, \tilde{k}_{AB}\,\tilde{k}^*_{CD} + \frac{16}{9} \bar{k}_{AB}\,\big(\bar{k}^*_{CD}+\bar{k}^*_{DC}\big)\Big],\nonumber\\
c_3 \big[A,B,C,D\big] &+& \bigg(\frac{\langle\sigma\rangle^2}{\Lambda^2\,M_T^2} + \frac{D\big(M^2_T,M^2_\sigma\big)}{\Lambda^2}\bigg)\,                           \Big[ 64\,h_{AB}\,h^*_{CD} + 36\, \bar{h}_{AB}\, \bar{h}^*_{CD} + \frac{128}{9}\, g_{AB}\,g_{CD} \Big],\nonumber\\
c_4 \big[A,B,C,D\big] &=& \bigg(\frac{\langle\sigma\rangle^2}{\Lambda^2\,M_T^2} + \frac{D\big(M^2_T,M^2_\sigma\big)}{\Lambda^2}\bigg)\,                            \big( 64\ h_{AB}\,\bar{h}_{CD}  \big).
\eeqa
where, $M_{T} $ and $M_{\sigma}$ are the masses of Triplet and $\sigma$ respectively. The expression inside the parenthesis of eq. (\ref{eq:p1}) is comprised of two terms: the first term corresponds to the contribution at tree-level, while the second term arises from the contribution at loop-level. In general, $\big\langle\sigma\big\rangle\neq M_{\sigma}$ and the exact relation between $\sigma$ and $M_{\sigma}$ depend on the contents of the scalar potential. However, it is assumed that $M_{\sigma} = \langle \sigma \rangle $ through this manuscript, implying mass of $B-L$ violating scalar is equal to its vev. Moreover, these effective strengths of the various operators depend on the Yukawa couplings and the mass of the mediators. It is evident from the above coefficients in eq. (\ref{eq:p1}), $\sigma-T$ can generate all the four operators mentioned in eq. (\ref{eq:bnc}). Despite $\sigma$ being charged under $B-L$, the resulting operators when $\sigma$ acquires vev conserve $B-L$, as is expected from the corollary that operators having $B-L=0$  typically appear at even mass dimensions \cite{Kobach:2016ami,Helset:2019eyc}.

The loop contribution in eq. (\ref{eq:p1}) depend upon the loop integral, whose definition is as follows.
\beqa \label{eq:lf}
D\big(M_1^2,M_2^2\big) &\equiv& -\frac{1}{16\pi^2}\Bigg( \frac{M_1^2\log\Big(\frac{M_1^2}{\mu^2}\Big)-M_2^2\log\Big(\frac{M_2^2}{\mu^2}\Big)}{M_1^2-M_2^2}-1\Bigg). 
\eeqa
The loop function given in the above expression in eq. (\ref{eq:lf}) is symmetric under the exchange of $M_1$ and $M_2$, where $M_1$ and $M_2$ are the masses of scalars propagating inside the loop. Furthermore, the renormalisation scale, $\mu$, can be chosen as $1$ GeV, corresponding to the scale at which the proton at rest can decay. Ideally, all parameters contributing to proton decay should be renormalised down to this scale ($1$ GeV).

Further, the next pair capable of having diquark and leptoquark couplings is $T-\Theta$ and the contribution to the effective $D=6$ operators upon its mediation is computed below. 
\beqa \label{eq:p2}
c_1 \big[A,B,C,D\big] &=& \frac{1}{\Lambda^2}\, D\big(M_T^2,M^2_\Theta\big)\,\big( -32\,h_{AB}\,h_{CD} \big)\nonumber\\
c_2 \big[A,B,C,D\big] &=& \frac{1}{\Lambda^2}D\big(M_T^2,M^2_\Theta\big)\,\Big[ -64\,h_{AB}\,\bar{h}^*_{CD} -8\,\tilde{h}_{AB}\,\tilde{h}^*_{CD} \Big] \nonumber\\
c_3 \big[A,B,C,D\big] &=& \frac{1}{\Lambda^2}D\big(M_T^2,M^2_\Theta\big)\,\Big[ \frac{-64}{9}\,\bar{g}_{AB}\,\bar{g}^*_{CD}  - 2\big(\tilde{k}_{AB}+ \frac{3}{10}\tilde{k}_{BA}\big)\, \big( \frac{1}{2}\tilde{k}^*_{CD}- \tilde{k}^*_{DC}\big)\nl &-& \,\frac{2}{3}\, \left(\bar{k}_{AB}-\frac{5}{4}\bar{k}_{BA}\right)\,\left(\bar{k}^*_{CD}-4\bar{k}_{DC}\right) \Big] \nl
c_4 \big[A,B,C,D\big] &=& \frac{1}{\Lambda^2}D\big(M_T^2,M^2_\Theta\big)\,\big( -\frac{64}{9}\, \bar{g}_{AB}\,g_{CD} \big) 
\eeqa
here, $M_{T},\,\text{and}\, M_{\Theta}$ are the masses of $T$ and $\Theta$ respectively. None of the scalars can assume a vev in this case, as it would violate $SU_3^{\rm{C}}\times U_1^{\rm{em}}$ symmetry. Consequently, this pair can only induce proton decay at the loop level and contribute to all four operators mentioned in eq. (\ref{eq:bnc}).

Next, we consider the case when the pair $T-H$ mediates the proton decay.
\beqa {\label{eq:p3}}
c_2[A,B,C,D] &=&  -64\,\bigg(\frac{\langle H\rangle^2}{\Lambda^2\,M_T^2} + \frac{D\big(M^2_H,M^2_T\big)}{\Lambda^2}\bigg)\,\big(\hat{h}_{AC}+\hat{h}^*_{CA}\big)\,\big(\hat{h}_{BD}+\hat{h}^*_{DB}\big)
\eeqa
$M_{H}$ being the mass of the $H$. While $H$ typically transforms as the SM Higgs, however, it is charged under $B-L$ (cf. Table (\ref{tab:tab1})), limiting $H$ from assuming a large vev. This pair can mediate tree and loop-level proton decays, contributing to effective $D=6$ operators. However, it can generate only a single $D=6$ effective operator, ${\mathcal{O}}_2$. Additionally, to arrive at the above expression of eq. (\ref{eq:p3}), Fierz transformation rules for two-component spinors have been used \cite{Dreiner:2008tw}.

Next, we consider the case when the pair $H-\Delta$ mediates the proton decay. Its contribution to effective $D=6$ operators is shown below.
\beqa \label{eq:p4}
c_1 \big[A,B,C,D\big] &=& \Big( \frac{\langle H\rangle^2}{\Lambda^2\,M^2_{\Delta}} + \frac{D\big(\mh,\md\big)}                   {\Lambda^2}\Big)\,\big( 32\,h_{AB}\,\bar{h}_{CD} -           \frac{64\sqrt{2}}{9}\,g_{AB}\,\bar{g}_{CD}   \big)\nonumber\\
c_2 \big[A,B,C,D\big] &=& \Big( \frac{\langle H\rangle^2}{\Lambda^2\,M^2_{\Delta}} + \frac{D\big(\mh,\md\big)}{\Lambda^2}\Big)\,\big( 16\,h_{AB}\,h^*_{CD} + \frac{16}{9} g_{AB}\,g^*_{CD} + 32\,\tilde{h}_{AB}\,\tilde{h}^*_{CD} \big) \nonumber\\
c_3 \big[A,B,C,D\big] &=& \Big( \frac{\langle H\rangle^2}{\Lambda^2\,M^2_{\Delta}} + \frac{D\big(\mh,\md\big)}{\Lambda^2}\Big)\,\Bigg[ 32\,\bar{h}_{AB}\,\bar{h}^*_{CD} + \frac{64}{9}\,\bar{g}_{AB}\,g^*_{CD} \big) \nl
                        &+&  \left(\tilde{k}_{AB}-2\sqrt{2}\tilde{k}_{BA}\right)\,\left( \left(\frac{1}{\sqrt{2}}-\frac{3}{20}\right)\tilde{k}^*_{CD} + \left(\sqrt{2}-\frac{3}{20}\right)\tilde{k}^*_{DC}\right)                         \nl
                        &- &  2\left(\frac{1}{\sqrt{2}}\bar{k}_{AB}-\frac{2\sqrt{2}}{3}\bar{k}_{BA}\right)\left( \sqrt{2}\big(k_{CD}+2 k_{DC}\big) - \big( \frac{\sqrt{2}}{48} \bar{k}_{DC}+\frac{4\sqrt{2}}{3}\bar{k}_{CD}\big)\right) \Bigg]\nl
c_4 \big[A,B,C,D\big] &=& \Big( \frac{\langle H\rangle^2}{\Lambda^2\,M^2_{\Delta}} + \frac{D\big(\mh,\md\big)}{\Lambda^2}\Big)\,\big( 32 \bar{h}_{AB}\,h_{CD} -\frac{32}{9}\,g_{AB}\,g_{CD} \big) 
\eeqa
$M_{H}$, and $M_{\Delta}$ being the respective masses of $H$ and $\Delta$. Analogous to the previous case, this pair also mediate proton decays at tree and loop levels and contributes to all four effective $D=6$ proton decay operators. 

Further, the pair $\Delta-\Delta$ induces only loop-level proton decay, as shown below.
\beqa \label{eq:p5}
c_1 \big[A,B,C,D\big] &=& \frac{1}{\Lambda^2}\, D\big(M_\Delta^2,M^2_\Delta\big)\,\big( - 64                                \bar{h}_{AB}\,h_{CD} + \frac{10}{9}\,\bar{g}_{AB}\,g_{CD}  \big)\nl
c_2 \big[A,B,C,D\big] &=& \frac{1}{\Lambda^2}D\big(M_\Delta^2,M^2_\Delta\big)\,\Big[                                       -16\sqrt{2}\,\bar{h}_{AB}\,\bar{h}^*_{CD} + \frac{10}{9}\, \bar{g}                              _{AB}\,g^*_{CD} + \frac{8}{405}\,f_{AB}\,f^*_{CD}\big) \nonumber\\
                       &+& \frac{36\sqrt{2}}{5}\, \tilde{k}_{AB}\,\tilde{k}^*_{DC}+64 \left(\frac{1}{\sqrt{2}}-\frac{16}{3}\right)\bar{k}_{AB}\,\bar{k}^*_{DC} \bigg]\nl
c_3 \big[A,B,C,D\big] &=& \frac{1}{\Lambda^2}D\big(M_\Delta^2,M^2_\Delta\big)\,\big( 64 h_{AB}                            \,h^*_{CD} + \frac{8}{3}\, g_{AB}\,\bar{g}_{CD} -8 g_{AB}\,g^*_{CD}+ \frac{8}{9}\,\tilde{g}_{AB}\,\tilde{g}^*_{CD}\big)\nonumber\\
c_4 \big[A,B,C,D\big] &=& \frac{8}{\Lambda^2}D\big(M_\Delta^2,M^2_\Delta\big)\,\big( 16 h_{AB}\,                           \bar{h}_{CD} - \frac{8}{3}\, g_{AB}\,\bar{g}_{CD} \big) 
\eeqa
This particular pair contributes to all the effective $D=6$ operators. 

The pair $\Delta-\Theta$ mediates proton decay at loop level and contributes to only operator $\mathcal{O}_2$.
\beqa {\label{eq:p6}}
c_2[A,B,C,D] &=&  \frac{D\big(M^2_{\Delta},M^2_{\Theta}\big)}{\Lambda^2}\,\Big[\big(\hat{h}_{AD}+\hat{h}^*_{DA}\big)\,\big(\hat{h}_{CB}+\hat{h}^*_{BC}\big) + \frac{16\sqrt{2}}{9}\hat{g}_{AC}\big(\hat{g}_{BD} + \hat{g}^*_{BD}\big) \nl &+& \frac{16\sqrt{2}}{45}\,\hat{f}_{AD}\,\hat{f}_{BC}\Big]
\eeqa

Similarly, the pair $\Delta-t$ mediates proton decay at loop only and leads to only operator ${\mathcal{O}_2}$. 
\beqa {\label{eq:p7}}
c_2[A,B,C,D] &=&  \frac{D\big(M^2_{\Delta},M^2_{t}\big)}{\Lambda^2}\,\Big[ \frac{8\sqrt{2}}{9}\hat{g}_{AC}\big(\hat{g}_{BD} + \hat{g}^*_{BD}\big) - \frac{8\sqrt{2}}{45}\,\hat{f}^*_{DA}\,\hat{f}^*_{BC}\Big]
\eeqa
where, $M_t$ is the mass of $t$.

Finally, contribution to operators, $\mathcal{O}_{1,2,3,4}$, is shown below when the proton decay is induced by the pair $\Theta-t$. 
\beqa \label{eq:p8}
c_1 \big[A,B,C,D\big] &=& \frac{1}{\Lambda^2}\, D\big(M_{\Theta}^2,M^2_t\big)\,\big(16 \bar{h}                              _{AB}\,h_{CD} + \frac{10}{9}\,\bar{g}_{AB}\,g_{CD}  \big)\nonumber\\
c_2 \big[A,B,C,D\big] &=& \frac{1}{\Lambda^2}D\big(M_{\Theta}^2,M^2_t\big)\,\bigg[ 16\,\bar{h}                               _{AB}\,\bar{h}^*_{CD} + \frac{10}{9}\, \bar{g}                                                _{AB}\,g^*_{CD} + \frac{8}{405}\,f_{AB}\,f^*_{CD}\big) \nonumber\\
                      &-&   \frac{8}{20}\ \tilde{k}_{AB}\,\big(\tilde{k}^*_{CD}+\tilde{k}^*_{DC}\big)- \frac{16}{15} \bar{k}_{AB}\,\left(\frac{29}{4}\,\bar{k}^*_{CD} + \bar{k}^*_{DC}\right)\bigg]\nl
c_3 \big[A,B,C,D\big] &=& \frac{1}{\Lambda^2}D\big(M_{\Theta}^2,M^2_t\big)\,\big( 64 h_{AB}                                 \,h^*_{CD}  + \frac{8}{3}\, g_{AB}\,\bar{g}^*_{CD} + 4 g_{AB}                                       \,g^*_{CD} - \frac{16}{9}\tilde{g}_{AB}\,\tilde{g}^*_{CD}\big)\nonumber\\
c_4 \big[A,B,C,D\big] &=& \frac{1}{\Lambda^2}D\big(M_{\Theta}^2,M^2_t\big)\,\big( 32 h_{AB}\,                               \bar{h}_{CD} - \frac{8}{3}\, g_{AB}\,\bar{g}_{CD} \big) 
\eeqa

The expressions derived in eqs. (\ref{eq:p1}, \ref{eq:p2}, \ref{eq:p3}, \ref{eq:p4}, \ref{eq:p5}, \ref{eq:p6}, \ref{eq:p7} and \ref{eq:p8}) quantifies the contribution of different pairs of scalars capable of mediating $B$ and $L$ violating but $B-L$ conserving nucleon decays. These coefficients depend on the masses of the scalars and determinable Yukawa couplings. The following points summarise the notable features of these evaluated coefficients.

\begin{itemize}
    
    \item Only three pairs of scalars can mediate nucleon decay at tree level, viz $\sigma-T$, $H-T$ and $H-\Delta$. $\sigma-T$ and $H-\Delta$ contribute to all the operators given in eq. (\ref{eq:bnc}), while $H-T$ contributes to only $\mathcal{O}_2$. The immediate consequence of the latter is that $H-T$ can only induce charged antilepton modes accompanied by neutral mesons, which are $p\to e^+ \pi^0$, $p\to \mu^+\,K^0$ and others.
    \item Other five pairs, $T-\Theta$, $\Delta-\Delta$, $\Delta-\Theta$, $\Delta-t$ and $\Theta-t$,  can induce nucleon decay only at the loop level. Of these pairs, $T-\Theta$, $\Delta-\Delta$, and $\Theta-t$ can contribute to all the four operators and induce proton decaying into charged and neutral anti-lepton while $\Delta-\Theta$ and $\Delta-t$ can contribute to $\mathcal{O}_2$ only.
    \item For the case of Yukawa couplings, terms with Yukawa coupling $\hat{h}$, $\hat{g}$ and $\hat{f}$ can only give rise to those terms which only contribute to $\mathcal{O}_2$ and hence offer a clean signature of detectability.
    \item Terms with Yukawa couplings $\tilde{h}$ and $\tilde{g}$ contribute to the operators $\mathcal{O}_{2}$ and $\mathcal{O}_{3}$.
    \item Expressions in which the intermediate integrated out irreps are $10_{H}$ and $120_{H}$ contribute to all four operators of eq. (\ref{eq:bnc}) while when the intermediate integrated out irrep are $45_{H}$ and $210_{H}$, contributes to operators $\mathcal{O}_2$ and $\mathcal{O}_3$. Additionally, the integration of $126^{\dagger}_{H}$ as an intermediate irrep yields only operator $\mathcal{O}_{2}$ of eq. (\ref{eq:bnc}).
    \item Term with Yukawa coupling $k$ does not contribute to nucleon decay at all as it only gives rise to either diquark or dilepton vertices.
    \item In renormalisable GUTs, only two scalars, along with their conjugate partners, are capable of mediating $B-L$ conserving tree-level proton decay modes with SM charges $\left(3,1,\frac{1}{3}\right)$ and $\left(3,3,-\frac{1}{3}\right)$. Additionally, $B-L$ violating modes at the tree level are mediated by the pair $\left(3,2,\frac{1}{6}\right)$ - $\left(3,1,\frac{1}{3}\right)$ along with its conjugate partner \cite{Patel:2022wya}. However, in the case of non-renormalisable GUTs, the lowest dimension at which tree-level $B-L$ conserving proton decay occurs is $D=8$, always mediated by a pair of scalars as discussed above.
\end{itemize}

Nevertheless, these pairs of scalars may also contribute to $B-L$ violating two body proton decays \cite{Babu:2012iv,Babu:2012vb,Patel:2022wya}. However, if the lowest dimension at which $B-L$ conserving proton decays appear is $D=8$, we anticipate that $B-L$ violating two-body proton decay modes would emerge at $D=9$ and beyond, specifically odd dimensions, and thus will be suppressed. Consequently, we constrain the mediators based on the $B-L$ conserving two-body proton decay modes. It is to be noted that there could be other sources of four fermion operators which can induce proton decays. As we aim to constrain the scalars residing in $16_{\mathrm{H}}$ through proton decays, such sources of four fermion operators have been neglected.

\section{Constraints from Proton Decay} {\label{sec:pds}}
So far, we have identified different pairs of scalars capable of inducing tree as well as loop-level proton decays and analysed their contribution, in terms of the Wilson coefficient, to the effective $D=6$ proton decay operators. The estimated coefficients depend on the explicit Yukawa couplings, also contributing to the masses of charged and neutral fermions. In the current scenario, these couplings appear at the non-renormalisable level in the Lagrangian.

The class of renormalisable \so\, theories featuring the Yukawa sector with complex  $10_{H}$ and $126^{\dagger}_{H}$ is known for reproducing the observed fermion mass spectrum and mixing angles \cite{Joshipura:2011nn,Mummidi:2021anm}, and the Lagrangian is shown below.
\beqa {\label{eq:realso10}}
-\mathcal{L}_{R} &=&  \fs_A\,\Big( H_{AB}\,10_H + F_{AB}\,126_{H}^{\dagger} \Big)\fs_B  
\eeqa 
$H$ and $F$ are the symmetric Yukawa couplings in the generation space. 

One can evade the coupling of $126^{\dagger}_{H}$ with $\fs$-plet fermion and instead include $16_H$ which couples to $\fs$-plet at non-renormalisable level. The Lagrangian of such a set-up is shown below.
\beqa{\label{eq:SO10RNR}}
-{\cal{L}}_{\rm{Y}} &=& H_{AB}\,\fs_A\,\fs_B\,10_H + \frac{y_{AB}}{\Lambda} \left(\fs_A\,\fs_B\,16_H\,16_H\right)
+\, \frac{1}{\Lambda} \bar{y}_{AB}\,\big(\mathbf{16}_{A}\,\mathbf{16}_{B}\;16^{\dagger}_{H}\,16^{\dagger}_{H}\big)\hc\nl
\eeqa

In the absence of $126^{\dagger}_{H}$, its role will be typically performed by $16_H$ for a realistic \so\, model. Further, the non-renormalisable term mentioned in eq. (\ref{eq:SO10RNR}) can be decomposed only via $10_{H}$-dimensional \so\, irrep as the coupling of different irreps with $\fs$-plet are denied. Such an absence can be ensured by introducing a discrete symmetry which only allows coupling of $16$-plet with $10_H$ at the renormalisable level and with $16_H$ at non-renormalisable level. Consequently, $h$
and $\bar{h}$ coupling mentioned in eq. (\ref{eq:16Fs16Hs}) are allowed, and they could presume the order of Yukawa coupling of $126_{H}$ with $\fs$-plet fermion.
\beqa {\label{eq:hNR}}
h\frac{\langle\sigma\rangle}{\Lambda} &= & \bar{h}\frac{\langle\sigma\rangle}{\Lambda} \hspace{0.5cm}\sim \hspace{0.5cm}  F.
\eeqa
Furthermore, we impose the condition that the entries of $h$ and $\bar{h}$ remain perturbative, i.e. $|h|$ and $|\bar{h}|$ $\leq 4\pi$. The largest entry of $F$ in a renormalisable $SO(10)$ based on $10_{H}$ and $\overline{126}_{H}$ is of the order of $10^{-3}$~(cf. \cite{Mummidi:2021anm}). The largest entry of $F$, together with the perturbativity requirement, places a lower bound on the parameter $\sigma$ as $10^{-3.5}\Lambda$, while $\sigma$ itself is not subject to an upper bound. We set $\Lambda$ as the reduced Planck scale ($\sim\,10^{18.5}$ GeV), making the $B-L$ scale closer to the typical GUT scale. Opting for a $B-L$ scale lower than the aforementioned scale renders the couplings  $f$ non-perturbative and thus making \so\, unrealistic. The best-fit value of the entries of $F$ can be inferred from \cite{Mummidi:2021anm}.

Further, once the \so\, gauge symmetry is broken into SM gauge symmetry, the following expressions of effective Yukawa couplings arise.
\beqa {\label{eq:yukrel}}
Y_{u,d,e,\nu} &=& c_1^{u,d,e,\nu} H + \frac{\langle\sigma\rangle}{\Lambda} \big(c_2^{d,\,e}\, h +  c_3^{u,\,\nu}\,\bar{h} \big), \nl
M_{R} &=&  r\frac{\langle\sigma\rangle^2}{\Lambda} \big( h + \bar{h}\big) . 
\eeqa

where, $c_{1}^{u,d,e,\nu}$ are the various ${\mathcal{O}}\big(1\big)$ Clebsch Gordan coefficients and can be inferred from \cite{Mummidi:2021anm}.  $c_{2,3}^{u,d,e,\nu}$ and $r$  are also the ${\mathcal{O}}\big(1\big)$ Clebsh Gordan coefficients.  When $\mathbf{\sigma}$,  $B-L$  charged singlet-scalar present in $16_H$, assumes vev, Majorana mass associated right-handed neutrinos is generated, alike the case of renormalisable \so. Further, light neutrino masses can be generated via invoking Type-I seesaw mechanism \cite{Minkowski:1977sc,Yanagida:1979as,Mohapatra:1979ia,Schechter:1980gr}.

\subsection{Branching Pattern}
\label{ssec:brpattern}
In our scenario, we consider the contribution from the non-renormalisable part of the Yukawa Lagrangian, as presented in eq. (\ref{eq:SO10RNR}), to proton decay to be comparable to the renormalisable part, and collectively both these contributions should adhere to the constraint imposed by proton decay. Consequently, the Yukawa couplings $h$ and $\bar{h}$ become the primary contributors to proton decay stemming from the non-renormalisable couplings. As it is evident from eq.(\ref{eq:p1}) only the pairs $\sigma-T$ and $H-\Delta$ can potentially induce proton decay at the tree level.   

The expressions of partial decay widths of leading proton decay modes are given in appendix \ref{eq:pdecay}. Table (\ref{tab:pdecayvariation}) shows the branching pattern of such proton decay modes, assuming the contribution of a pair of scalars capable of inducing proton decay at the tree level.  

%%%%%%%%%%%%%
\begin{table}[t]
\begin{center}
\begin{tabular}{lccc} 
\hline
Branching ratio [\%]& ~~~$\sigma-T$~~~  & ~~~$H-\Delta$~~~ & ~~~$\sigma-T\,+ H-\Delta $~~~\\
& $M_\Delta\to\infty$ & $M_{T} \to \infty$  &~~~with $M_T=M_{\Delta}$\\
\hline\hline
${\rm BR}[p\to e^+ \pi^0]$ & $< 1$ & $<1$ & $< 1$\\
${\rm BR}[p\to \mu^+ \pi^0]$  & $2$ & $ 2$ & $2$\\
${\rm BR}[p\to \bar{\nu} \pi^+]$  & $12$ & $13$ & $13$\\
${\rm BR}[p\to e^+ K^0]$ & $< 1$ & $< 1$ & $< 1$\\
${\rm BR}[p\to \mu^+ K^0 ]$  & $3$ & $2$ & $2$\\
${\rm BR}[p\to \bar{\nu} K^+]$  & $83$ & $84$ & $82$\\
${\rm BR}[p\to e^+ \eta]$ & $<1$  &  $<1$ & $<1$\\
${\rm BR}[p\to \mu^+ \eta]$  & $<1$ &  $<1$ & $< 1$\\
\hline
\end{tabular}
\end{center}
\caption{Proton decay branching fractions estimated for different hierarchies, assuming $\big\langle \sigma\big\rangle = 10^{14}$ GeV and $\big\langle H \big\rangle = 246$ GeV.}
\label{tab:pdecayvariation}
\end{table}
%%%%%%%%%%%%%%%%%%%%

The crucial observations regarding Table (\ref{tab:pdecayvariation}) are as follows.
\begin{enumerate} [(a)]
    \item In the case of scalar-mediated proton decay, the proton favours decay into the second-generation mesons compared to the first ones. This is due to the hierarchical nature of Yukawa couplings, which has also been noticed in the case of renormalisable $SO(10)$ models \cite{Patel:2022wya}. {\label{pt:aa}}
  \end{enumerate}

The above-mentioned feature of the proton decay spectrum can be understood from the flavour structure. The realistic fermion mass spectrum in the underlying model leads to the following hierarchical structure of $h$ and $\bar{h}$ from which matrices that diagonalises $Y_{u,d,e}$ could be obtained and are given as follows \cite{Mummidi:2021anm}.

\be \label{eq:HF_form}
f = \bar{f} \sim \frac{\Lambda}{\sigma}\, F \sim \frac{\Lambda}{\sigma}\, \lambda^4\, \left(\ba{ccc} \lambda^5 & \lambda^4 & \lambda^3\\
\lambda^4 & \lambda^3 & \lambda^2\\
\lambda^3 & \lambda^2 & \lambda \ea\right),\; \text{and}\; U_{u,d,e} \sim\left(\ba{ccc} 1 & \lambda & \lambda^3\\
\lambda & 1 & \lambda^2 \\
\lambda^3 & \lambda^2 & 1 \ea\right),~~U_\nu \sim \left( \ba{ccc} &  &  \\  & {\cal O}(1) & \\  &  &  \ea \right)\,\ee
where $\lambda = 0.23$ is Cabibbo angle. From eq. (\ref{eq:HF_form}), we find 
\beqa \label{eq:UFU}
~~U_f^T\,f\,U_{f^\prime} \sim  \frac{\Lambda}{\sigma} \,\lambda^4\, \left(\ba{ccc} \lambda^5 & \lambda^4 & \lambda^3\\ \lambda^4 & \lambda^3 & \lambda^2\\
\lambda^3 & \lambda^2 & \lambda \ea\right)\,=U_d^T\,f\,U_\nu \eeqa
where $f,f^\prime = u,d,e$. Substitution of the above results of eq. (\ref{eq:UFU}) in the expressions of partial decay widths of the proton as mentioned in the eq. (\ref{eq:pdecay}), one finds the following. The definition of various parameters entering in the following ratios are also defined in the appendix (\ref{sec:a1}). 

\beqa \label{eq:ppattern}
\frac{\Gamma[p \to \overline{\nu}\,\pi^+]}{\Gamma[p \to \overline{\nu}\,K^+]} & \simeq & \frac{(m_p^2 - m{^2_{\pi^+}})^2}{(m_p^2 - m{^2_{K^+}})^2}\,\frac{(1+\tilde{D}+\tilde{F})^2}{\left(1+\frac{m_N}{m_S}(\tilde{D}-\tilde{F})+\frac{m_N}{m_{\Lambda}}\big(\tilde{D}+3\tilde{F}\big)\right)^2}\,  {\lambda}^2 \simeq 2 \lambda^2\,.\nl
\frac{\Gamma[p \to e^+\,\pi^0]}{\Gamma[p \to \mu^+\,\pi^0]} &\simeq& \frac{\Gamma[p \to e^+\,K^0]}{\Gamma[p \to \mu^+\,K^0]} \simeq \lambda^2\,, \nl
\frac{\Gamma[p \to \mu^+\,K^0]}{\Gamma[p \to \overline{\nu}\,\pi^+]}  &\simeq & \frac{(m_p^2 - m{^2_{K^0}})^2}{(m_p^2 - m{^2_{\pi^+}})^2}\,\frac{\frac{1}{2}\left(\frac{m_N 64}{m_S }(\tilde{D}-\tilde{F})\right)^2+ \frac{1}{2}64^2}{(1+\tilde{D}+\tilde{F})^2\,32^2}\,   \simeq 0.2 \,,\nl
\frac{\Gamma[p \to \mu^+\,K^0]}{\Gamma[p \to \overline{\nu}\,K^+]}  &\simeq & \frac{(m_p^2 - m{^2_{K^0}})^2}{(m_p^2 - m{^2_{K^+}})^2}\,\frac{\frac{1}{2}\left(\frac{m_N 64}{m_S }(\tilde{D}-\tilde{F})\right)^2+ \frac{1}{2}64^2}{\left(1+\frac{m_N}{m_S}(\tilde{D}-\tilde{F})+\frac{m_N}{m_{\Lambda}}\big(\tilde{D}+3\tilde{F}\big)\right)^2}\,   \simeq 0.2\lambda^2 \,.
\eeqa

The expressions given in eq.(\ref{eq:ppattern}) explains the claim mentioned in point (a) and explain the branching pattern of the proton given in the table (\ref{tab:pdecayvariation}) mediated by $\sigma-T$. Further, the branching pattern of proton decay induced by $H-T$ yields similar ratios. 

\begin{enumerate}[(b)]
    \item  {\label{pt:bb}} In the context of a realistic \so\ scenario, the precise values of the $B-L$ scale and $\Lambda$ are irrelevant for the tree-level branching pattern of the proton in the case of $\sigma-T$. 
\end{enumerate}
The aforementioned observation can be understood as follows. As indicated by eq. (\ref{eq:p1}), the tree-level contribution of the coefficient of a specific pair effectively scales as $c\sim \frac{\langle\sigma\rangle^2}{\Lambda^2} \big|h\big|^2$, where $h$ represents a particular Yukawa coupling. As deduced from eq. (\ref{eq:hNR}), for a realistic \so\, scenario, $h\sim \frac{\Lambda}{\langle\sigma\rangle}$, rendering the coefficient seemingly independent of $\big\langle\sigma\big\rangle$ and $\Lambda$. The nature of $h$ is dictated by the requirement to reproduce the observed fermion mass spectrum. However, opting for an alternative value of $h$ independent of $\sigma$ and $\Lambda$ introduces a dependence of the coefficient on the latter, potentially lowering down the mass of the mediator.

\begin{enumerate}[(c)]
    \item  The loop contribution of the pair of scalars to the proton decay is highly suppressed when the $B-L$ scale is larger than $10^{11}$ GeV, and hence, the loop contribution has been ignored while computing the branching pattern of the proton. \label{p:c}
\end{enumerate}
    
Claim (c) can be understood considering the loop contribution of $\sigma-T$ to the proton decay. Substituting eqs. (\ref{eq:p1}) and (\ref{eq:UFU}) into the partial decay width of proton decay, as provided in eq. (\ref{eq:pdecay}), resulting in the decay into $\nu\,K^+$, yields the following expression.

\beqa{\label{eq:ptc}}
\Gamma[p\to \overline{\nu}\,K^+] &=& \frac{\big(D\big(M_T^2,M_{\sigma}^2\big)\big)^2}{\big\langle\sigma\big\rangle^4} \frac{(m_p^2 - m_{K^\pm}^2)^2}{32\, \pi\, m_p^3 f_\pi^2} \left(1+\frac{m_N}{m_S}(\tilde{D}-\tilde{F})+\frac{m_N}{m_{\Lambda}}\big(\tilde{D}+3\tilde{F}\big)\right)^2\,32^2\,\lambda^{30}\nl
\eeqa
where different symbols have their usual meaning provided in the appendix (\ref{sec:a1}) and $D\big(M_T^2,M_{\sigma}^2\big)$ is the loop function defined in the eq. (\ref{eq:lf}). The expression given in  the above eq. (\ref{eq:ptc}) can modified to yield $\sigma$ as follows.
\beqa {\label{eq:ptcc}}
\big\langle\sigma\big\rangle &=& \Bigg[\tau_{p\to \overline{\nu}\,K^+}\frac{\big(D\big(M_T^2,M_{\sigma}^2\big)\big)^2}{\big\langle\sigma\big\rangle^4} \frac{(m_p^2 - m_{K^\pm}^2)^2}{32\, \pi\, m_p^3 f_\pi^2}\nl
& &\left(1+\frac{m_N}{m_S}(\tilde{D}-\tilde{F})+\frac{m_N}{m_{\Lambda}}\big(\tilde{D}+3\tilde{F}\big)\right)^2\,32^2\,\lambda^{30}\Bigg]^{0.25} \nonumber
\eeqa
Using the current lower bound on the lifetime of $p\to \nu\,K^+$ \cite{Super-Kamiokande:2014otb}, we get the following. 
\beqa{\label{eq:ptccc}}
\big\langle \sigma \big \rangle &>& \Big[D\big(M^2_T,M^2_{\sigma}\big)\Big]^{0.5}\, 2.5*10^{11}\; \rm{GeV}
\eeqa
The loop function provided above reaches its maximum value when both masses are degenerate, and its value is always less than 1, considering the fact that the maximum mass the scalar can attain is $M_p$. Consequently, $\sigma$ must be greater than $10^{11}$ GeV to satisfy the proton decay bound. In realistic \so\ scenarios, $\sigma$ typically lies around $10^{14}$ GeV, and lowering $\sigma$ below than $10^{14}$ GeV hampers the perturbativity of Yukawa couplings.  As a result, the pair that mediates proton decays at the loop level cannot be strictly constrained in this scenario.

\subsection{Estimation of bound on the mediator}
Having obtained the branching pattern of leading proton decay modes by the pair of scalars able to induce tree-level proton decay modes, we constrain their masses using the current lower bound on the lifetime of proton \cite{Super-Kamiokande:2014otb}.

The lower bound on the mass of pair $\sigma-T$ from its tree-level contribution is as follows.
\beqa \label{eq:limit_T}
\tau/{\rm BR}[p \to \overline{\nu}\, K^+] &=& 5.9 \times 10^{33}\,{\rm yrs}\,\times \left(\frac{M_{T}}{2.8 \times 10^{11}\,{\rm GeV}} \right)^4\,.\eeqa

Similarly, the bound on the mass of $\Delta$ from its tree-level contribution is the following.
\beqa \label{eq:limit_D}
\tau/{\rm BR}[p \to \overline{\nu}\, K^+] &=& 5.9 \times 10^{33}\,{\rm yrs}\,\times \left(\frac{M_{\Delta}\,\big\langle\sigma\big\rangle}{4.9 \times 10^{13}\,{\rm GeV}^2} \right)^4\,\left(\frac{ 246 }{\big\langle H\big\rangle}\right)^4.\eeqa

Eqs. (\ref{eq:limit_T}) and (\ref{eq:limit_D}) quantify the bounds on $M_T$ and $M_\Delta$ respectively. The bound on $M_T$ is independent of the $B-L$ scale, while the bound on $\Delta$ depends on the ratio of $\frac{\langle\sigma\rangle}{\langle H\rangle}$. Both mediators, $T$ and $\Delta$, remain independent of the precise value of the cutoff scale ($\Lambda$) due to the consideration of a realistic \so\ scenario. Moreover, the mass of $\Delta$ relies on the ratio $\frac{\langle\sigma\rangle}{\langle H\rangle}$, which can be adjusted to achieve a lighter mass for $\Delta$, preferably at the TeV scale. This possibility raises interesting prospects for probing $\Delta$ in future colliders \cite{Dorsner:2016wpm,Raj:2016aky,Padhan:2019dcp}. Additionally, in renormalisable SO(10) models, $\Delta$ also mediates $B-L$ violating proton decay modes. The bound on its mass from $p\to \nu\,K^+$ turns out to be greater than $10^6$ GeV \cite{Patel:2022wya}. Thus, the mass bound on the $\Delta$ gets significantly reduced in our scenario.

\section{Summary and Discussion} {\label{sec:conc}}

Proton decay is one of the crucial predictions of GUTs and is an important yardstick to constrain the new particles (scalar and vector), which are inextricable parts of the GUTs. Its detection would discard the baryon number conservation principle and ultimately provide hope to unify fundamental interactions. 

This study investigates a version of the non-renormalisable $SO(10)$ model that lacks large tensor representations. The $16_{H}$ representation is the smallest irrep, which couples with the $\fs$-plet, generating charged and neutral fermion masses at the non-renormalisable level. We begin with decomposing the coupling of $16_{H}$ with the $\fs$-fermions, exploring all possible decompositions allowed by integrating out different  $SO(10)$ irreps, into $SU(5)$ invariant expressions. After that, we calculate the couplings of scalar pairs with SM fermions stemming from these $SU(5)$ decomposed expressions capable of inducing proton decay. Subsequently, we assess the tree and loop level contributions of different pairs of scalars to the effective $D=6$ $B-L$ conserving operators. We also argued that the $16_{H}$ irrep could fulfil the role typically played by $126^{\dagger}_{H}$ in renormalisable \so\, GUTs, thus assigning its Yukawa coupling to be similar to that of $126^{\dagger}_{H}$. Finally, we impose constraints on these pairs of scalars residing in the $16_{H}$ representation through proton decays. The main features of this study are summarised as follows.

\begin{itemize}
    \item The proton decay in the non-renormalisable $SO(10)$ is always mediated by  pairs of scalars and can happen in two distinct topologies, resulting into effective $D=6$ operator. Firstly, at the tree level, one scalar is integrated out while the other acquires a vev, or vice versa. Secondly, both scalars contribute to proton decay through loops. 
\item In total, there are eight different pairs capable of inducing proton decay. Out of them, only three can mediate proton decay at the tree level, $\sigma-T$, $H-\Delta$ and $H-T$, while other pairs induce proton decay at the loop level.

\item In a non-renormalisable $SO(10)$ model with $10_{H}$ and $16_{H}$ participating in the Yukawa sector, proton decay at the tree level is mediated by the pair $\sigma-T$ and $H-\Delta$. Due to the hierarchical Yukawa couplings, proton favours to decay in the second generation mesons accompanied with charges or neutral antileptons. For a realistic scenario, the lower bound on $M_T$ is greater than $10^{11}$ GeV, while $M_\Delta$ is constrained to be at least 1 TeV. The bound on $M_\Delta$ arises from collider searches.

\end{itemize} 

In the renormalisable version of $SO(10)$ GUTs, proton decay at the lowest mass dimension, i.e. six, is mediated by a single scalar particle, and these mediator can typically remain lighter than the conventional GUT scale. However, in the case of the non-renormalisable version, proton decay at the lowest dimension is always mediated by a pair of scalars. In some scenarios, these mediators can remain closer to the electroweak scale than the GUT scale, thus can also account for the lower-energy associated phenomenology. Non-renormalisable GUT models offer compelling solutions to the inconsistencies of renormalisable GUT models in the UV regime and may allow a promising avenue for the unification of gravitational interactions with the known fundamental interactions.

\section*{Acknowledgements}
I express my gratitude to Namit Mahajan and Ketan M. Patel for valuable discussions, suggestions, encouragement, and meticulous review of the manuscript. 

\appendix

\section{Proton Decay Width Expression}
{\label{sec:a1}}
The decay width expressions of leading proton decay modes can be computed using the chiral perturbation theory \cite{Claudson:1981gh,Chadha:1983sj} and are given as follows \cite{Nath:2006ut,Beneito:2023xbk}.

\beqa{\label{eq:pdecay}}
\Gamma[p \to e_i^+\pi^0] &=& \frac{(m_p^2 - m_{\pi^0}^2)^2}{32\, \pi\, m_p^3 f_\pi^2} A^2 \left( \frac{1+\tilde{D}+\tilde{F}}{\sqrt{2}}\right)^2  \Big(\left| \alpha\, y[u_1,d_1,e^C_i,u^C_1] + \beta\, y^{\prime *}[u^C_1,d^C_1,e^C_i,u^C_1] \right|^2 \Big. \nonumber \\
& + & \Big. \left| \alpha\, y^*[u^C_1,d^C_1,e_i,u_1] + \beta\, y^\prime[u_1,d_1,e_i,u_1]\right|^2 \Big), \nonumber\\
\Gamma[p \to \overline{\nu}\pi^+] &=& \frac{(m_p^2 - m_{\pi^\pm}^2)^2}{32\, \pi\, m_p^3 f_\pi^2} A^2 \left(1+\tilde{D}+\tilde{F} \right)^2 \sum_{i=1}^3 \left| \alpha\, y^*[u^C_1,d^C_1,\nu_i, d_1] + \beta\, y^\prime[u_1,d_1,\nu_i, d_1] \right|^2\,, \nonumber\\
\Gamma[p \to e_i^+K^0] &=& \frac{(m_p^2 - m_{K^0}^2)^2}{32\, \pi\, m_p^3 f_\pi^2} A^2\, \frac{1}{2} \Big[\Big|C_{Li}^- - C_{Ri}^- +\frac{m_p}{m_B}(\tilde{D}-\tilde{F})\left(C_{Li}^+ - C_{Ri}^+\right)\Big|^2 \Big. \nonumber \\
& + & \Big. \Big|C_{Li}^- + C_{Ri}^- +\frac{m_p}{m_B}(\tilde{D}-\tilde{F})\left(C_{Li}^+ + C_{Ri}^+\right)\Big|^2 \Big]\,, \nonumber\\
\Gamma[p \to \overline{\nu}K^+] &=& \frac{(m_p^2 - m_{K^\pm}^2)^2}{32\, \pi\, m_p^3 f_\pi^2}\, A^2\, \sum_{i=1}^3 \Big| \frac{2\tilde{D}}{3} \frac{m_p}{m_B} C^\nu_{L1i} + \left(1+\frac{\tilde{D}+3\tilde{F}}{3} \frac{m_p}{m_B}\right) C^\nu_{L2i} \Big|^2, \nonumber\\
\Gamma[p \to e_i^+\eta] &=& \frac{(m_p^2 - m_{\eta}^2)^2}{32\, \pi\, m_p^3 f_\pi^2}\, A^2\, \frac{1}{6}\Big[ \Big|C_{L i}^+ (1-\tilde{D}+3\tilde{F}) - 2 C_{L i}^-\Big|^2 \, \nonumber\\
& + & \Big. \Big| C_{R i}^+ (1-\tilde{D}+3\tilde{F}) - 2 C_{R i}^-\Big|^2\Big]\, \Big.,
\eeqa
where,
\beqa \label{eq:C_dw}
C^\pm_{L i} &=& \alpha\, c_3^*[u^C_1,d^C_2,u_1,e_i] \pm \beta\, c_1[u_1,d_2,u_1,e_i]\,,
\nonumber\\
C^\pm_{R i} &=& \alpha\, c_2^*[u_1,d_2,u^C_1,e^C_i] \pm \beta\, c_4[u^C_1,d^C_2,u^C_1,e^C_i]\,,
\nonumber\\
C^\nu_{L1i} &=& -\alpha\, c_3^*[u_1^C,d_2^C,d_1,\nu_i] - \beta\, c_1[u_1,d_2,d_1,\nu_i]\,,\nonumber \\
C^\nu_{L2i} &=& -\alpha\, c_3^*[u_1^C,d_1^C,d_2] - \beta\, c_1[u_1,d_1,d_2,\nu_i]\,. \eeqa
The various $C's$ defined in the eq. (\ref{eq:C_dw}) are in the flavour basis and should be converted appropriately into the mass basis. Also, $m_H$ represents the mass of various hadrons denoted by $H$, including protons ($p$), neutral pions ($\pi^0$), neutral kaons ($K^0$), charged kaons ($K^\pm$). $m_B$ signifies the average mass of baryons, while $f_\pi$ denotes the pion-decay constant. Parameters $\alpha$, $\beta$, $\tilde{D}$, and $\tilde{F}$ are the input chiral Lagrangian parameter. The parameter $A$ accounts for renormalization effects in hadronic matrix elements, particularly from the weak scale to the proton mass, $m_p$. The numerical values to these parameters in eq. (\ref{eq:pdecay}): $\alpha = 0.01,{\rm GeV}^3$, $\tilde{D} = 0.8$, and $\tilde{F} = 0.46$ \cite{JLQCD:1999dld, Cabibbo:2003cu}. Furthermore, we utilize the average baryon mass, $m_B = 1.15$ GeV, and the pion decay constant, $f_\pi = 130$ MeV, alongside various hadron masses sourced from the Particle Data Group (PDG) \cite{Zyla:2020zbs}. The parameter $A$ is set to $1.43$ to accommodate running effects from the electroweak scale ($M_Z$) to the proton mass scale. Although typically, values of $F$ obtained at the electroweak scale are prone to RGE effects from the GUT scale to $M_Z$, we opt for their GUT-scale values due to minimal changes in Yukawa couplings during running, which shall have negligible impacts on the results \cite{Alonso:2014zka}.

\bibliography{references}

\end{document}